\documentclass[12pt,a4paper,oneside]{article}
\usepackage{graphicx,amssymb,amsmath,color}
\usepackage[linkcolor={blue},citecolor={red},colorlinks=true]{hyperref}
\usepackage{cite}
\usepackage{relsize}
\usepackage{enumerate}
\usepackage[margin=3 cm]{geometry}
\usepackage{slashed}
\usepackage{soul}
\usepackage{multirow}
\usepackage[normalem]{ulem}
\usepackage[dvipsnames, x11names]{xcolor}
\usepackage[compat=1.0.0]{tikz-feynman}
\usepackage{xcolor}
\usepackage{mathtools}

\newcommand{\be}{\begin{equation}}
\newcommand{\ee}{\end{equation}}

\def\beq{\begin{equation}}
\def\eeq{\end{equation}}
\def\bea{\begin{eqnarray}}
\def\eea{\end{eqnarray}}

\def\bit{\begin{itemize}}
\def\eit{\end{itemize}}

\def\ra{\rightarrow}

\def\baa{\begin{array}}
\def\eaa{\end{array}}

\def\simgt{\mathrel{\lower2.5pt\vbox{\lineskip=0pt\baselineskip=0pt
           \hbox{$>$}\hbox{$\sim$}}}}
\def\simlt{\mathrel{\lower2.5pt\vbox{\lineskip=0pt\baselineskip=0pt
           \hbox{$<$}\hbox{$\sim$}}}}

\def\bfc{\begin{figure}\begin{center}}
\def\efc{\end{center}\end{figure}}

\definecolor{chromeyellow}{rgb}{1.0, 0.65, 0.0}
\definecolor{darkcoral}{rgb}{0.8, 0.36, 0.27}
\definecolor{cadmiumgreen}{rgb}{0.0, 0.42, 0.24}

\begin{document}

 \begin{flushright}
DESY-24-110
\end{flushright}
\vspace{.6cm}
\begin{center}

{\Large \bf 
Leptogenesis via Bubble Collisions\\}

\vspace{1cm}{Martina Cataldi$^{1,2}$ and Bibhushan Shakya$^1$}
\\[7mm]

{$^1$ \it Deutsches Elektronen-Synchrotron DESY, Notkestr.\,85, 22607 Hamburg, Germany}\\

{$^2$ \it II. Institute of Theoretical Physics, Universität Hamburg, , Luruper Chaussee 149, 22761, Hamburg, Germany}
\end{center}

\bigskip \bigskip \bigskip

\centerline{\bf Abstract} 
\begin{quote}
We present a novel realization of leptogenesis from the decays of sterile (right-handed) neutrinos (RHNs) produced from runaway bubble collisions at a first order phase transition. Such configurations can produce heavy RHNs with mass many orders of magnitude above the scale of symmetry breaking as well as the temperature of the plasma, thereby enabling high scale leptogenesis without the need for high reheat temperatures while also naturally suppressing washout effects. This mechanism is also efficient for RHN masses $\gtrsim 10^{14}$ GeV, the natural scale for type-I seesaw with $\mathcal{O}(1)$ couplings, where standard thermal leptogenesis faces strong suppression from washout processes in equilibrium. The corresponding phase transitions are at scales $\gtrsim\!10^9$ GeV and produce gravitational wave signals that could be detected by future experiments. 

\end{quote}

\newpage

\tableofcontents

\section{Introduction}

The origin of the observed baryon asymmetry of the Universe (BAU) is one of the most striking puzzles of fundamental particle physics and early Universe cosmology. Many mechanisms for the dynamical generation of the asymmetry have been proposed in various beyond the Standard Model (BSM) contexts (see e.g.~\cite{Elor:2022hpa,Barrow:2022gsu} for recent reviews). The necessary ingredients for such mechanisms to be viable are captured by the three Sakharov conditions: (i) baryon number violation, (ii) $CP$ violation, and (iii) out-of-equilibrium condition.

An elegant mechanism that can produce the observed BAU while drawing motivation from other shortcomings of the Standard Model (SM) of particle physics is thermal leptogenesis\cite{FUKUGITA198645}. An extension of the SM by sterile / right-handed Majorana neutrinos (RHNs) via the terms 
\be
\mathcal{L}\supset y_{\nu(ij)}\, \overline{L}_i h N_j+M_{N (j)} \overline{N}^c_j N_j\,,
\ee
where $L_i$ are the SM lepton doublets and $h$ is the SM Higgs doublet, is known to produce SM neutrino masses $m_\nu\approx y_{\nu}^2 v_h^2/M_N$, where $v_h =246$ GeV is the SM Higgs vev, via the type-I see-saw mechanism\cite{Minkowski:1977sc, PhysRevLett.44.912, Schechter:1980gr, Schechter:1981cv}. The smallness of SM neutrino masses indicates a high scale for the RHN masses: e.g. for $\mathcal{O}(1)$ values of $y_{\nu}$, one requires $M_N\sim 10^{14}$ GeV. If such RHNs are produced from the thermal bath in the early Universe, their decays can satisfy the Sakharov conditions for the generation of the baryon asymmetry: $CP$ violation can be realized in the above Yukawa interactions, the out-of-equilibrium condition is naturally realized by RHN decays, whereas baryon number violation is achieved through electroweak sphaleron processes, which can convert the lepton asymmetry generated due to the Majorana nature of the RHNs into a baryon asymmetry. This thermal leptogenesis mechanism is known to successfully produce the observed baryon asymmetry of the Universe for  $10^{9}$ GeV $\lesssim M_N \lesssim 10^{13}$ GeV in the one-flavor approximation (for some leptogenesis reviews, see \cite{Buchm_ller_2005,Davidson_2008, B_deker_2009}).  Below the lower bound $M_N\approx 10^{9}$ GeV, the corresponding Yukawa couplings are too small to generate the observed asymmetry in the absence of resonant effects, while beyond upper bound  $M_N \approx 10^{13}$ GeV, the $\Delta L =2$ washout process $ L \, h \to \bar{L} \, h $ remains in equilibrium when the RHNs decay, which can erase the produced asymmetry, hence requiring a more careful treatment of various aspects in the calculation (see discussions in \cite{Giudice:2003jh}). The window of viability can be extended with resonant effects with degenerate masses
(see \cite{Pilaftsis:2003gt, Granelli:2020ysj, Pilaftsis:2005rv} for reviews), or with nonthermal production of the RHNs \cite{Chung:1998rq,Giudice:1999fb,Hahn-Woernle:2008tsk, Dasgupta:2022isg,Zhang:2023oyo,Bhandari:2023wit,Tong:2024lmi,Cataldi:2024bcs,Azatov:2021irb, Huang:2022vkf, Chun:2023ezg, Dichtl:2023xqd}.

Standard thermal leptogenesis, while a simple and elegant mechanism, features several undesirable aspects. First, the mechanism is only successful within a small window of RHN masses; in particular, the RHN mass scale $M_N\sim 10^{14}$ GeV most natural from the point of view of type-I seesaw with $\mathcal{O}(1)$ Yukawa couplings is in the regime of strong washout. Second, a high reheat temperature above the RHN mass scale is required. Third, there is no experimentally testable aspect of leptogenesis at such high scales. In this paper we propose a novel framework for leptogenesis that improves on all of these aspects. 

Our mechanism makes use of runaway bubble collisions at a first order phase transition (FOPT). FOPTs \cite{Hogan:1983ixn, Witten1984, Hogan:1986qda, KosowskyTurnerWatkins-GW-92, kosowsky92-1, Kamionkowski_1994} are generally predicted in many BSM extensions, and have recently been extensively studied in the literature as a promising cosmological source of gravitational waves (GWs) \cite{Grojean:2006bp,Caprini:2015zlo,Caprini:2018mtu,Caprini:2019egz,Athron:2023xlk}. If the bubbles of true vacuum
are not slowed down by friction effects and achieve the so-called runaway behavior, the collisions of such runaway bubbles can produce extremely heavy particles much heavier than the scale of the phase transition or the temperature of the plasma \cite{Watkins:1991zt,Konstandin:2011ds,Falkowski:2012fb,Mansour:2023fwj,Shakya:2023kjf,Giudice:2024tcp}. This provides a unique setting for producing ultra-heavy RHNs without reheating the Universe to such high temperatures. Since the masses of the RHNs produced from bubble collisions can be several orders of magnitude above the temperature of the bath following the FOPT, thermal washout processes that erase the produced lepton asymmetry are also naturally suppressed. In particular, we will see that leptogenesis with $M_N\sim 10^{14}$ GeV now also becomes viable without washout concerns. Finally, the sizable GWs produced by such FOPTs also provide a testable aspect of this leptogenesis mechanism. 

The idea of using $CP$-violating, out-of-equilibrium decays of heavy particles to populate the BAU has been explored in various contexts over the years (see e.g. \cite{Dimopoulos:1987rk,Claudson:1983js,Sorbello:2013xwa,Cui:2012jh,Arcadi:2015ffa,Grojean:2018fus,Pierce:2019ozl}). The prospects of producing such heavy particles from bubble collisions at a FOPT were examined in \cite{Katz_2016}. More recently, several papers \cite{Azatov:2021irb,Baldes:2021vyz, Baldes_2021, Chun:2023ezg} have investigated baryogenesis/leptogenesis scenarios 
making use of heavy particle production during bubble expansion due to the sudden mass gain of particles when they cross into bubbles of true vacuum, which enables them to up-scatter into heavy states. Our framework differs from the setups in \cite{Azatov_2021, Baldes_2021, Chun:2023ezg} in two main aspects. First, we do not require the RHNs to gain mass from the FOPT, which allows us to consider more generic dark FOPTs unrelated to the breaking of baryon and/or lepton number. Second, the RHN masses can consequently be many orders of magnitude heavier than the scale of symmetry breaking or the temperature of the thermal bath, for which runaway bubble collisions, which we consider, provide the more efficient production mechanism compared to the bubble-plasma interactions studied in previous works. These differences carry two main practical implications: high scale leptogenesis is possible with a reheat temperature that can be several orders of magnitude lower, and leptogenesis at the natural type-I seesaw scale, $M_N\gtrsim 10^{14}$ GeV, is readily realized. 

We first discuss various aspects of the framework used for our study, such as the FOPT configurations and parameters relevant for our study and the formalism to calculate particle production from bubble collisions, in Section \ref{sec:framework}.
In order to illustrate the broad applicability of the idea and the qualitative differences between different scenarios, we explore leptogenesis from bubble collisions in two different setups: one where the background field undergoing the FOPT couples directly to the heavy RHNs (the neutrino portal setup, Section \ref{sec:nplepto}), and another where they couple indirectly through a scalar field (the scalar portal setup, Section \ref{sec:splepto}). We summarize the main points of our paper in Section \ref{sec:summary}.

\section{Framework}
\label{sec:framework}

In this section we describe the framework of our study: the relevant FOPT parameters and configurations, the formalism for calculating particle production from bubble collisions, the basics of leptogenesis, and the production of gravitational waves from FOPTs. 

\subsection{FOPT Parameters}
\label{parameters}

We consider a FOPT involving a background scalar field $\phi$ transitioning from a false / metastable vacuum charaterized by vacuum expectation value (vev) $\langle \phi \rangle=0$ to a true /stable vacuum characterized by $\langle \phi \rangle=v_\phi$ through the nucleation, expansion, collision and percolation of true vacuum bubbles.  The difference in the potential energies of the two vacua provides the latent energy released in the phase transition, which we parameterize as 
\be
\Delta V \equiv V_{\langle \phi \rangle=0}-V_{\langle \phi \rangle=v_\phi}=c_V\,v_\phi^4\,.
\ee

In addition, the following phenomenological FOPT parameters will be relevant to our study:

\begin{itemize}
\item $\alpha\equiv\frac{\rho(\text{vacuum})}{\rho(\text{radiation})}$: the strength of the phase transition, the ratio between $\rho(\text{vacuum})=\Delta V$ and the total energy density in the radiation bath at the onset of the phase transition. 

\item $\beta$: (inverse) duration of phase transition, characterized by a dimensionless parameter $\beta/H$, where $H$ is the Hubble scale.

\item $v_w$: velocity of the bubble wall, which tends to grow as the bubble walls gain the energy released in the phase transition and accelerate.  

\item $\gamma_w=1/\sqrt{1-v_w^2}$: the Lorentz boost factor of the bubble wall. We are interested in the ultrarelativistic regime $v_w\approx 1,~\gamma_w\gg 1$. 

\item $l_w$: bubble wall thickness in the plasma frame, which Lorentz contracts as $l_w=l_{w_0}/\gamma_w$, where $l_{w_0}\sim\mathcal{O}(v_\phi^{-1})$ is the wall thickness at bubble nucleation.

\item $T_n$: the temperature of the thermal bath when true vacuum bubbles begin to nucleate at a rate greater than the Hubble scale, triggering the FOPT. 

\item $R_0$: bubble radius at nucleation, typically $\mathcal{O}(T_n^{-1})$. 

\item $R_*$: the typical size of vacuum bubbles at collision, determined from the timescale over which the transition completes, $R_*\approx v_w\,(8\pi)^{1/3}{\beta}^{-1}$. 

\item $T_*$: the temperature of the thermal bath at the end of the phase transition. Generally, $T_*\approx T_n$ unless the Universe becomes vacuum dominated during the transition.
\end{itemize}

In this paper, we take a model-independent approach and do not specify any underlying theory for the FOPT, thus treating the above parameters as independent. Consequently, our results can be easily applied to any specific model, where the above quantities can be calculated from the fundamental parameters of the underlying model. 

\subsection{Runaway Phase Transitions}
\label{subsec:runaway}

In order to produce RHNs with mass many orders of magnitude higher than the scale of the phase transition as well as the thermal bath temperature, we are interested in FOPT configurations that achieve runaway behavior, where bubble walls accelerate to large boost factors and become energetic enough to produce such heavy particles at collision. In general, a FOPT can occur either due to thermal effects, where temperature-dependent corrections cause the true vacuum to emerge or become energetically favored, or purely via quantum tunneling, which can occur in the absence of a plasma. In the presence of a thermal plasma, there are two friction contributions that can prevent runaway behavior. The leading order pressure comes from particles crossing into the bubbles and becoming massive 
\cite{Bodeker:2009qy,Dorsch:2018pat,Espinosa:2010hh}:
\be
\mathcal{P}_{\text{LO}}\approx \frac{1}{24}m^2 T^2 \,,
\label{eq:pressure}
\ee
where $m$ is the mass of the particle in the broken phase and $T$ is the temperature. For relativistic walls, the next-to-leading order contribution, corresponding to radiation of gauge bosons, produces pressure that scales approximately as \cite{Bodeker:2017cim,Hoche:2020ysm,Gouttenoire:2021kjv,Ai:2023suz,Long:2024sqg}
\be
\mathcal{P}_{\text{NLO}}\sim g^2\, \gamma_w\, m_V\, T^3 \,,
\label{eq:pressure2}
\ee
where $m_V$ is the mass of the gauge boson and $g$ is the gauge coupling. The walls continue to accelerate and become more energetic as long as $\Delta V$ exceeds these two contributions. 
Such $\gamma_w\gg1$ runaway behavior can therefore be realized in several distinct ways (see \cite{Giudice:2024tcp} for a more detailed discussion):
 
\begin{itemize}
\item Thermal transition without a gauge boson, with $\Delta V>\mathcal{P}_{\text{LO}}$.

\item Thermal transition with a light gauge boson $(m_V\ll v_\phi)$: the frictional term $\mathcal{P}_{\text{NLO}}$ (Eq.\,\ref{eq:pressure2}) becomes relevant at large boost factors, and the wall boost factor at collision is 
$\gamma_w\sim \text{min}\left[\frac{c_V}{g^3},~\frac{2 R_*}{3 R_0}\right]\,$, which can be large since the gauge coupling is small ($g\ll1$).

\item Supercooled phase transition:  In a supercooled FOPT \cite{Konstandin:2011dr,vonHarling:2017yew,Ellis:2018mja,Baratella:2018pxi,DelleRose:2019pgi,Fujikura:2019oyi,Ellis:2019oqb,Brdar:2019qut,Baldes:2020kam}, $\Delta V > \rho_\text{radiation}$, diluting the pre-existing thermal bath due to a brief vacuum-dominated period. The walls therefore encounter negligible friction and attain runaway behavior. In this case, reheating at the end of the phase transition creates a thermal bath with $\rho_\text{radiation}=\Delta V$. 

\item Quantum tunneling in a cold sector: The sector associated with the FOPT can be cold, and the transition can occur purely due to quantum tunneling without any thermal effects. In this case, a vacuum dominated epoch can be avoided by assuming that there is comparable or greater energy density in the form of a SM radiation bath. 

\end{itemize}

In this paper, we assume that the bubble walls achieve runaway behavior, but remain agnostic about which of the above scenarios gives rise to the configuration. In the runaway regime, the boost factor of the bubble wall grows as $\gamma\approx \frac{2 R}{3 R_0}$ \cite{Ellis:2020nnr} and can reach extremely large values at collision. Using the parameters and relations defined above, and assuming $R_0\approx 10/v_\phi$ at nucleation, particle production is therefore possible up to the maximum energy
\beq
E_{\text{max}}=\gamma_{\text{max}}/l_{w0}\sim 0.07 \left(\frac{\alpha}{(1+\alpha)\,c_V}\right)^{1/2}\frac{M_{Pl}}{\beta/H}\,.
\label{emax}
\eeq
We will make use of this possibility to produce heavy RHNs from bubble collisions.

\subsection{Particle Production from Bubble Collisions}
\label{sec:particle_prod} 

Particle production can take place during various stages of a FOPT; production during collision dominates over production during the nucleation and expansion phases for particles heavier than the scale of the phase transition \cite{Shakya:2023kjf,Giudice:2024tcp}. The collision of bubbles was first considered in \cite{Hawking:1982ga}, and the formalism to calculate particle production from such collisions was first presented in \cite{Watkins:1991zt}. Based on this formalism, analytic results were derived in simplified ideal limits in \cite{Falkowski:2012fb}, and recently refined with numerical studies of more realistic setups in \cite{Mansour:2023fwj} and analytic treatment in \cite{Shakya:2023kjf}, with issues of gauge dependence discussed in \cite{Giudice:2024tcp} . Here we provide a brief outline of the formalism; the interested reader is referred to \cite{Watkins:1991zt,Falkowski:2012fb,Mansour:2023fwj,Shakya:2023kjf,Giudice:2024tcp} for greater details.

In this formalism, the background scalar field configuration is treated as a classical external background field and the states coupled to it as quantum fields in the presence of this source. Particle production is computed by calculating the imaginary part of the effective action of the background field configuration. For this purpose, the classical background field configuration is decomposed via Fourier transform into modes of definite four-momenta $p^2=\omega^2-p_z^2>0$, which are interpreted as off-shell propagating field quanta of the background field with effective mass $m^2=p^2$. These mode excitations, denoted by $\phi^*_p$, can decay into quanta of all fields that $\phi$ couples to. The number of particles thus produced per unit area of bubble collision (assuming planar bubble walls) can be written as 
\cite{Watkins:1991zt,Falkowski:2012fb}
\be
\frac{N}{A}= 2 \int\frac{dp_z\,d\omega}{(2\pi)^2}\,|\tilde{\phi}(p_z,\omega)|^2 \,\text{Im}[\tilde{\Gamma}^{(2)}(\omega^2-p_z^2)] \,,
\label{interpretation}
\ee
where $\tilde{\phi}$ is the Fourier transform of the field configuration and $\tilde{\Gamma}$ is the two point 1PI Green function. 
It is convenient to reformulate this as
\cite{Falkowski:2012fb} 
 \be
\frac{N}{A}=\frac{1}{2 \pi^2}\int_{p_{\text{min}}^2}^{p_{\text{max}}^2} d p^2\,f(p^2) \,\text{Im} [\tilde{\Gamma}^{(2)}(p^2)] \, .
\label{number}
\ee
The lower limit of the integral is given by either the mass of the particle species being produced, i.e. $p_{\text{min}}=2m$ for pair production, or by the inverse size of the bubble, $p_{\text{min}}=(2R_*)^{-1}$, since at lower momenta the existence of multiple bubbles needs to be taken into account. The upper limit is given by the maximum energy available in the collision process, $p_{\text{max}}=2/l_w=2\gamma_w/l_{w0}$. $f(p^2)$ represents the \textit{efficiency factor} for particle production at a given energy scale $p$ and encapsulates the details and nature of the collision. From numerical studies of realistic bubble collision processes, the efficiency factor can be parameterized as \cite{Mansour:2023fwj} 
\be
f_{\text{elastic}}(p^2)= f_{\mathrm{PE}}(p^2)+\frac{v_{\phi}^2L_p^2}{15 m_{\mathrm{t}}^2}\exp{\left(\frac{-(p^2 - m_{\mathrm{t}}^2+12 m_{\mathrm{t}}/L_p)^2}{440 \, m_{\mathrm{t}}^2 / L_p^2}\right)}\qquad \text{(elastic collisions)}
\label{eq:elasticfit}
\ee
\be
f_{\text{inelastic}}(p^2)= f_{\mathrm{PE}}(p^2)+\frac{v_{\phi}^2L_p^2}{4 m_{\mathrm{f}}^2}\exp{\left(\frac{-(p^2 - m_{\mathrm{f}}^2+31 m_{\mathrm{f}}/L_p)^2}{650 \, m_{\mathrm{f}}^2 / L_p^2}\right)}\qquad \text{(inelastic collisions)}
\label{eq:inelasticfit}
\ee
with $m_t,\,m_f$  the scalar masses in the true and false vacua respectively,  $L_p=\text{min}(R_*, \Gamma_\phi^{-1})$, where $\Gamma_\phi$ is the decay rate of the scalar as it performs oscillations around its true or false minimum. The efficiency factor for a perfectly elastic collision has been derived analytically in \cite{Falkowski:2012fb}
\be
f_{\mathrm{PE}}(p^2)=\frac{16 v_{\phi}^2}{p^4}\, \text{Log}\left[\frac{2(1/l_w)^2-p^2+2(1/l_w)\sqrt{(1/l_w)^2-p^2}}{p^2}\right]\,.
\label{eq:felastic}
\ee

We are interested in cases with $M_N\gg m_t,m_f$, hence the second terms in Eqs.\,\ref{eq:elasticfit} and \ref{eq:inelasticfit} are irrelevant, and the RHNs are produced via $f_{PE}$. The produced number density is thus independent of whether the collisions are elastic or inelastic.

The final factor in Eq.\,\ref{number} encodes the particle physics information and gives the decay probability of the $\phi^*_p$ excitations into particles. The imaginary part of the 2-point 1PI Green function $\Gamma^{(2)}$ can be written, using the Optical theorem, as  \cite{Watkins:1991zt,Falkowski:2012fb}
\be
\text{Im} [\tilde{\Gamma}^{(2)}(p^2)]=\frac{1}{2}\sum_k \int d\Pi_k |\bar{\mathcal{M}}(\phi^*_p\to k)|^2\,,
\label{optical}
\ee
where the sum runs over all possible final states $k$ that can be produced from the background field excitations $\phi^*_p$, $|\bar{\mathcal{M}}(\phi_p^*\to k)|^2$ is the spin-averaged squared amplitude for the decay of $\phi^*_p$ into the given final state $k$, and $d\Pi_k$ is the relativistically invariant n-body phase space element. 

For decay into fermions via a Yukawa coupling $y_f \phi \chi_ f\bar{\chi}_f$, the above gives
\be
\text{Im} [\tilde{\Gamma}^{(2)}(p^2)]_{\phi^*_p\to \chi_ f\bar{\chi}_f}=\frac{y_f^2}{8\pi} p^2 (1-4m_{\chi_f}^2/p^2)^{3/2}\,\Theta(p^2-4m_{\chi_f}^2)\,.
\label{fermion}
\ee

Moreover, the scalar $\phi$ particles themselves can be produced through these background field excitations via the quartic term $\frac{\lambda_\phi}{4!}\phi^4$ in the scalar potential; this gives rise to the 2-body decay
\be
\text{Im} [\tilde{\Gamma}^{(2)}(p^2)]_{\phi^*_p\to\phi\phi}=\frac{\lambda_\phi^2 \,v_\phi^2}{8\pi}  (1-4m_\phi^2/p^2)^{1/2}\,\Theta(p-2m_\phi)
\label{2scalar}
\ee
as well as the 3-body process
\be
\text{Im} [\tilde{\Gamma}^{(2)}(p^2)]_{\phi^*_p\to3\phi}=\frac{\lambda_\phi^2 \,p^2}{3072\,\pi^3}  (1-9m_\phi^2/p^2)^{1/2}\,\Theta(p-3m_\phi)\,.
\label{3scalar}
\ee
Note that the 3-body process is suppressed by a loop factor due to an additional particle in the final state, but is proportional to $p^2$ rather than $v_\phi^2$ and thus can become more important at higher $p^2$ as it can be realized even in the $v_\phi\to 0$ limit where the symmetry is unbroken. Production of other scalar particles (e.g. that couple as $\frac{\lambda_s}{4} \phi^2\chi_s^2$) through two- and three-body decays can be analogously calculated.

Since the particles produced on the bubble wall collision surface (Eq.\,\ref{number}) diffuse out over the entire volume occupied by the bubble, the final number density of particles is given by
 \be
n=\frac{3}{4 \pi^2 R_*}\int_{p_{\text{min}}^2}^{p_{\text{max}}^2} d p^2\,f(p^2) \,\text{Im} [\tilde{\Gamma}^{(2)}(p^2)] \, .
\label{number2}
\ee
Likewise, the energy density in particles per unit area is 
 \be
\frac{E}{A}=\frac{1}{2\pi^2}\int_{p_{\text{min}}^2}^{p_{\text{max}}^2} d p^2\,p\,f(p^2) \,\text{Im} [\tilde{\Gamma}^{(2)}(p^2)] \, ,
\label{energy}
\ee
which also gets diluted over the volume of the bubble. 

We will make use of the above formulae to calculate the number density and energy distribution of RHNs produced from bubble collisions. 

\subsection{Leptogenesis}

In this section, we describe the calculation of the lepton and subsequently baryon asymmetry from the decays of the RHNs.
As in standard thermal leptogenesis, the RHNs undergo decays, inverse decays and 2-to-2 scatterings, which can involve top quarks and gauge bosons in the presence of a SM thermal bath, as discussed in detail in several extensive reviews of leptogenesis \cite{Buchm_ller_2005,Davidson_2008, B_deker_2009}. For concreteness, in this paper we assume a nearly degenerate mass spectrum for the RHNs, close to the scale $M_N$. While all RHNs can participate in the production of the lepton asymmetry, we will only consider a single RHN for simplicity, which should give an order of magnitude estimate of the produced asymmetry in the absence of fine-tuned effects. We will also assume that there is negligible lepton or baryon asymmetry in the Universe at the onset of the phase transition, which would be the case, e.g. if the Universe did not get reheated to temperatures above $M_N$. 

The goal is to reproduce the observed baryon asymmetry of the Universe 
\bea
   \label{eq:BAU}
    Y_{ B} \equiv \frac{n_{B}-n_{\overline{B}}}{s} \bigg|_{0} = (8.69 \pm 0.22) \times 10^{-11}\,,
\eea
where $n_{B}$, $n_{\overline{B}}$ and $s$ are the number densities of baryons, antibaryons and entropy at present time. A convenient and simple parameterization of the baryon asymmetry produced in our novel leptogenesis scenario is
\begin{equation}
\label{eq:BA_estimate}
    Y_B^{bubble} = Y_N^{BC} \, \epsilon_{CP} \, c_{sph} \, \kappa_{wash} \, \kappa_{new} \, .
\end{equation}
Here $Y_N^{BC}=n_N/s$ is the RHN yield produced from bubble collisions, $c_{sph}=28/79$ is the sphaleron conversion factor, and $\kappa_{wash}$ and $\kappa_{new}$ are efficiency factors parameretizing the washout effects from the standard processes in thermal leptogenesis and new ones that appear in our framework, respectively. Finally, $ \epsilon_{CP}$ is the $CP$-asymmetry parameter
\beq
 \epsilon_{CP} \equiv \frac{\Gamma(N\to L h) - \Gamma(N\to  \overline{L}\, \overline{h})}{\Gamma(N\to L h) + \Gamma(N\to  \overline{L}\, \overline{h})} 
    \,.
\eeq
This is known to simplify to $\epsilon_{CP}\approx y_\nu^2/(8\pi)$, with $y_{\nu}^2 \equiv \sum _{i=1}^3 \big((y_{\nu}) ^{\dagger} y_{\nu}\big)_{ii} = \frac{2 M_N}{v_h^2} m_{\nu_3}^{atm}$ (with $m_{\nu_3}^{atm}\approx 0.05$ eV), which is a lower bound for a degenerate RHN mass spectrum \cite{Davidson_2002, Chun:2023ezg}. 

The lepton asymmetry is generated in RHN decays $ N \to L \, h$ via interference between tree-level and one-loop diagrams. The RHN decay rate into $Lh$ is 
\beq
    \Gamma_{N\to Lh} = \frac{y_{\nu}^2 M_N}{8 \pi E_N/M_N} \,.
    \label{RHNdecay}
\eeq
Note that when the RHN is relativistic, $E_N\gg M_N$, the decay rate gets suppressed by the Lorentz boost (or time dilation) factor $\gamma_{N} \approx E_N/M_N$.

If a thermal SM bath is present while the RHNs are in existence, a lepton asymmetry can also be produced via $\Delta L =1$ scatterings with top quarks and gauge bosons. In this paper, we will use approximate expressions for all scattering cross sections, neglecting thermal masses and dropping subleading logarithmic pieces, and the reader is referred to \cite{ Giudice_2004, Pilaftsis_2004} for the complete expressions. We have checked numerically that such effects are subleading for $M_N>T_*$, which is the parameter space of primary interest for this paper. The dominant contributions come from two processes:
\begin{itemize}
\item $N \, t \rightarrow L \, Q_3$ and $N \, Q_3\to L \, t$. These interactions are mediated by a t-channel Higgs, and the corresponding interaction rate can be written as 
\beq
\Gamma_{t} \simeq  \left(\frac{T_n}{T_*}\right)^3 n_{th} ( \sigma_{Nt \to Q_3 L}+\sigma_{N Q_3 \to t L})\approx \left(\frac{T_n}{T_*}\right)^3 n_{th}  \frac{3 y_\nu^2 y_t^2}{2\pi (M_N^2 + 4 E_N T_*)} \,.
\ee
Here $n_{th} = g_t T_*^3/\pi^2$ is the thermal abundance, with $g_t=2$ for top quarks, and  the $(T_n/T_*)^3$ prefactor represents the entropy dilution of the pre-existing thermal bath from the energy released in the phase transition.
\item $N \, A\to  h \, L$ and $N \, h\to  A \, L$. These processes are mediated via SM neutrinos/Higgs boson in the t-channel, and the rate can be written as \cite{ Giudice_2004, Pilaftsis_2004}
\beq
\label{eq:gamma_bosonscat}
    \Gamma_{N \, A \ra L \,  h}  \simeq \left(\frac{T_n}{T_*}\right)^3 n_{th}(\sigma_{N A \to h L}+\sigma_{N h \to A L}) \approx \left(\frac{T_n}{T_*}\right)^3 n_{th}\frac{9 y_\nu^2 g_w^2}{16\pi (M_N^2 + 4 E_N T_*)}
\eeq
\end{itemize}

We use $y_t =1$ and $g_w = 0.65$ for the Yukawa and gauge couplings; they are known to deviate from these values by $\mathcal{O}(1)$ factors at high temperatures due to running effects \cite{Giudice_2004}, which we do not incorporate in our calculations. 

The $CP$-asymmetry parameter for scatterings with top quarks and SM gauge bosons is approximately the same as for RHN decays, $ \epsilon_{CP}^{S}\approx \epsilon_{CP}^{D}$, for $T < M_N$ \cite{Nardi_2007_asymmetry,Fong:2010bh}. Due to this, each RHN produces the same amount of asymmetry whether it decays or scatters, hence the final asymmetry is largely independent of such details.

There also exist washout processes that tend to erase the lepton asymmetry produced from the RHN decays and scatterings. A SM thermal bath, even if absent at the time of the FOPT, gets produced from the thermalization of the energy released in the transition, hence these washout effects are always present regardless of the nature of the phase transition.\footnote{In contrast, the SM thermal bath produced from the energy released in the phase transition might not be relevant for the \textit{generation} of the asymmetry if the thermalization timescale is slower than the decay rate of the RHNs, as is generally the case.} The standard washout processes due to SM particles, represented by the factor $k_{wash}$ in Eq.\,\eqref{eq:BA_estimate}, are \cite{Davidson_2008, Chun:2023ezg}
\begin{itemize}
    \item $L \, h \rightarrow N$\\
   This inverse decay rate is $\Gamma_{inv}= \frac{y_{\nu}^2 M_N}{8 \pi} e^{-M_N/T_*}$. For $M_N \gg T_* $, this rate is exponentially suppressed and thus irrelevant. 
    
    \item $Q_3 \, L \rightarrow N \, t$ and $t \, L \rightarrow N \, Q_3$  (similar for $~h \, L \to N \, A$, $~A \, L \to N \, h$) 
    
The rate for this process is 
    \beq
\Gamma_{ (Q_3 L \to N t , \, t L \to N Q_3)} \approx  n_{th}  \frac{3 y_\nu^2 y_t^2}{2\pi M_N^2} e^{-M_N/T_*}\,.
\ee
  In general, $\Gamma(\text{scattering})/\Gamma(\text{decay})\ll 1 $ (see Appendices in \cite{Davidson_2008, Chun:2023ezg}). This can be understood by noting that these processes are also Boltzmann suppressed due to the $N$ in the final state, but include an additional vertex term and propagator compared to the decay process above. 
    
    \item $N \, L \rightarrow Q_3 \, t$ via s-channel Higgs (similar for $N \, L \to h \, A$). 
    
    The rate for this process can be written as \cite{ Giudice_2004, Pilaftsis_2004} 
    \beq
    \Gamma_{N \, L \ra Q_3 \, t}  \simeq \left(\frac{T_n}{T_*}\right)^3  n_{th}\frac{3 y_\nu^2 y_t^2}{4\pi (M_N^2 + 4 E_N T_*)} \, .
    \eeq
     Unlike the processes above, note that this process shuts off when the RHNs have decayed and is therefore only relevant if there is a pre-existing SM bath at the time of the transition. 

    \item $h \, L \rightarrow \bar L \, h,~L\, L\to h\, h$\\
The rates of these $\Delta L =2$ scattering processes, mediated by the RHN in the s- and t-channels, can be written as 
\beq
    \Gamma_{h \, L \rightarrow \bar L \, h, \, L\, L\to h\, h}\approx n_{th} \frac{y_{\nu}^4} {8 \pi M_N^2}\,.
\eeq
 Note that, unlike the above processes, these processes do not involve a RHN in the initial or final state and are therefore not Boltzmann suppressed.
\end{itemize}

In addition to these standard SM processes, there also exist new washout processes within our FOPT framework. These are represented by the $k_{new}$ factor in Eq.\,\ref{eq:BA_estimate}, and will be discussed below in the relevant sections.

\subsection{Gravitational Waves}
\label{subsec:gw}

Here we provide a brief description of the gravitational wave (GW) signal expected to be produced from the FOPT, which provides a testable phenomenon related to the leptogenesis process of interest. 

FOPTs produce GWs through the scalar field energy densities in the bubble walls after collision\,\cite{Kosowsky:1991ua,Kosowsky:1992rz,Kosowsky:1992vn,Kamionkowski:1993fg,Caprini:2007xq,Huber:2008hg,Jinno:2016vai,Jinno:2017fby,Konstandin:2017sat,Cutting:2018tjt,Cutting:2020nla,Lewicki:2020azd}, through the production of sound waves\,\cite{Hindmarsh:2013xza,Hindmarsh:2015qta,Hindmarsh:2017gnf,Cutting:2019zws,Hindmarsh:2016lnk,Hindmarsh:2019phv} and turbulence\,\cite{Kamionkowski:1993fg,Caprini:2009yp,Brandenburg:2017neh,Cutting:2019zws,RoperPol:2019wvy,Dahl:2021wyk,Auclair:2022jod} in the surrounding plasma, or through nontrivial spatial configurations of feebly-interacting particles that gain energy through the FOPT\,\cite{Jinno:2022fom}. For runaway bubble configurations, GWs are primarily sourced by bubble wall collisions, i.e.\,the scalar field. The peak frequency of such a signal today, as obtained from the study in \cite{Cutting:2020nla}, can be written as \cite{Giudice:2024tcp,Baldes:2022oev}
\be
f_\text{peak(GW)}=15~\mu\text{Hz}~\frac{\beta}{H}\,g_*^{1/6}\left(\frac{T_*}{10^3\,\text{GeV}}\right)=20~\mu\text{Hz}~\frac{\beta/H}{g_*^{1/12}}\left(\frac{(1+\alpha)}{ \alpha}c_V\right)^{1/4}\left(\frac{v_\phi}{10^3\,\text{GeV}}\right),
\ee
which enables us to translate the scale of the phase transition $v_\phi$ to the peak GW frequency today. The GW spectrum can be written as \cite{Cutting:2020nla}
\beq
h^2 \Omega_{\text{GW}}(f) \equiv h^2 \frac{d \Omega_{\text{GW}}}{d \log(f)}=2 \times 10^{-6} \bigg(\frac{\alpha}{1+\alpha}\bigg)^2 \frac{S_{\phi}(f)}{g_*^{1/3} (\beta /H)^2} \, ,
\eeq
where the shape of the spectrum can be parameterized as 
\beq
S_{\phi}(f) = \frac{(a+b) f_\text{peak(GW)}^b f^a }{b f_\text{peak(GW)}^{(a+b)}+ a f^{(a+b)}}\,,
\eeq
with $a=0.742$ and $b=2.16$ \cite{Cutting:2020nla}. For frequencies below
\beq
f_* = 16~\mu\text{Hz}~\frac{\beta/H}{g_*^{1/12}}\left(\frac{(1+\alpha)}{ \alpha}c_V\right)^{1/4}\left(\frac{v_\phi}{10^3\,\text{GeV}}\right)\, ,
\eeq
the GW spectrum scales as $\Omega_{\text{GW}} \propto f^3$ for initially super-horizon IR modes \cite{RuthDurrer_2003, PhysRevD.79.083519, Barenboim:2016mjm, PhysRevD.102.083528,Hook:2020phx}.

\section{Neutrino Portal}
\label{sec:nplepto}

We will now study two qualitatively distinct particle physics setups to relate the RHNs to the background field undergoing the FOPT. In this section we consider a neutrino portal setup, where the RHNs couple directly to $\phi$. This will be followed by a scalar portal setup (Section \ref{sec:splepto}), where the coupling between the two is mediated by a heavy scalar. As we will see, the two setups will feature important qualitative differences. 

We assume that there are multiple RHNs, but with non-hierarchical masses, i.e.\,they all lie at the common mass scale $M_N$. As stated earlier, we will only consider leptogenesis from a single RHN for simplicity.  
We will also assume that $\phi$ couples to the SM through some coupling that enables $\phi$ particles to decay into SM states, e.g. via $|\phi|^2 |h|^2$, so that all released energy is eventually converted to the SM bath.

\subsection{Setup}

For the neutrino portal setup, the idea is to mirror the RHN-SM coupling in the hidden sector. To this end, in addition to the scalar $\phi$, consider a dark sector fermion $\chi$ such that $\phi\,\chi$ is a gauge singlet under the symmetry broken by the $\phi$ vev (analogous to the $Lh$ combination in the SM). This enables us to write the following Lagrangian terms for the RHN (see e.g.\,\cite{Roland:2014vba,Shakya:2018qzg,Roland:2016gli,Shakya:2016oxf,Roland:2015yoa,Shakya:2015xnx,Morrison:2022zwk} for details and various applications of such neutrino portal frameworks): 
\beq
\mathcal{L}\supset y_\nu \overline{L} h N+y_D \chi \phi N +M_N \overline{N}^c N\,.
\eeq
The first term is the standard SM Dirac neutrino mass, whereas the second term is its dark sector analogue that couples the RHN to dark sector states.  We are primarily interested in cases where the RHN mass is significantly higher than the scale of the phase transition and the temperature of the bath, $M_N\gg v_\phi, T_*$.  This implies that lepton number breaking is unrelated to the FOPT under consideration and occurs at much higher scales. 

Note that the fermion $\chi$ is the dark sector analog of a SM neutrino; it is massless in the unbroken phase, and gains a small mass $m_\chi\approx y_D^2\,v_\phi^2/M_N$ after the FOPT via the type-I seesaw mechanism in the same manner that the SM neutrinos become massive after electroweak symmetry breaking (EWSB).  The $\phi$ vev also results in a small mixing angle between $\chi$ and $N$; after EWSB, a small mixing between $\chi$ and the SM neutrinos is also generated.\,\footnote{A single RHN cannot give mass to both $\chi$ and a SM neutrino $\nu$, as one can write a linear combination between $\chi$ and $\nu$ with vanishing coupling to $N$, which remains a massless eigenstate. However, as mentioned earlier, we focus on a single RHN for simplicity, whereas in the full theory we assume there are multiple RHNs, which can account for the masses of all SM neutrinos as well as $\chi$.} The mixing angles are parametrically
\beq
\text{sin}\,\theta_{\chi N}= \frac{y_D v_\phi}{M_N},~~~~~\text{sin}\,\theta_{\chi\nu}=\frac{y_{\nu}\, v_h}{y_D\, v_\phi}\,.
\eeq
By virtue of this mixing angle, $\chi$ behaves like a light sterile neutrino. In particular, it can decay as $\chi\to L \, h$. In principle, this decay channel can contribute to the lepton asymmetry; in practice, the associated effective coupling, which feeds into the $CP$-asymmetry parameter, is sufficiently small that this contribution turns out to be negligible. Furthermore, for sufficiently small $y_D$ and/or $v_\phi$, $\chi$ can be light and long-lived, which can be problematic for cosmology. As we will see below, this turns out to be irrelevant in the parameter space of interest. Therefore, the existence of $\chi$ will be of no phenomenological relevance for the leptogenesis mechanism that is the focus of this paper, and it solely serves the purpose of enabling a direct interaction term between $\phi$ and $N$.

\subsection{RHN Abundance}

In this setup, the background field excitations can decay directly into the RHN as $\phi^*\to \chi \, N$. Using the formulae provided in Sec.\,\ref{sec:particle_prod}, the number density of RHNs from such decays can be calculated to be approximately
\beq
\label{eq:neu_numbdens_neu}
n_N\approx 1.6\,y_D^2\, \frac{\beta}{H}\left(\frac{30(1+\alpha)c_V}{\pi^2\alpha}\right)^{1/2} \frac{v_\phi^4}{M_{Pl}}\text{ln}\left(\frac{2 E_{\text{max}}}{M_N}\right)\,,
\eeq
which gives
\beq
\label{eq:neu_Y_neu}
Y_N=\frac{n_N}{s}\approx 4 y_D^2\, \frac{\beta}{H}\left(\frac{\pi^2\alpha}{30(1+\alpha)g_* c_V}\right)^{1/4} \frac{v_\phi}{M_{Pl}}\,\text{ln}\left(\frac{2 E_{\text{max}}}{M_N}\right)\,,
\eeq
where $E_{\text{max}}$ is the maximum energy of the RHN, as given in\,Eq.\,\ref{emax}. It can be checked that the RHNs are mostly produced at energies close to this value, i.e. typically $E_N\approx E_{\text{max}}$.

Likewise, the number density of $\phi$ particles produced from bubble collisions is 
\beq
\label{eq:neu_numbdens_scal}
n_\phi\approx 0.64\,\lambda_\phi^2 \frac{\beta}{H}\left(\frac{30(1+\alpha)c_V}{\pi^2\alpha}\right)^{1/2} \frac{v_\phi^4}{M_{Pl}}\,,
\eeq
where $\lambda_\phi$ is the scalar quartic coupling and we have assumed $m_\phi\approx v_\phi$. This corresponds to the contribution from two-body decays of the background field excitation $\phi^*\to \phi\phi$, which provides the dominant contribution to the scalar number density, and the scalar particles are produced with energy $E_\phi\approx m_\phi$. 

It is also interesting to consider the high energy component of the $\phi$ population produced from $\phi^*\to 3 \phi$ with $E_\phi > M_N$, for which
\beq
\label{eq:neu_numbdens_scal_high}
n_\phi (E_\phi > M_N) \approx \frac{\lambda_\phi^2}{48 \pi^2} \frac{\beta}{H}\left(\frac{30(1+\alpha)c_V}{\pi^2\alpha}\right)^{1/2} \frac{v_\phi^4}{M_{Pl}}\text{ln}\left(\frac{2 E_{\text{max}}}{3 M_N}\right)\,.
\eeq
Note that this is smaller than the $n_\phi$ from the two-body decay process by approximately a loop factor, but can be relevant for high energy processes. 

There is also a freeze-in contribution to RHN abundance due to the inverse decay process $\chi \, \phi \to N$ once the dark sector bath is produced at $T=T_*$. This contribution is approximately
\beq
\label{eq:fi}
    n_N^{FI}\sim \frac{\Gamma(\chi \, \phi \to N)}{H(T_*)}n_{th} 
\eeq
with $\Gamma(\chi \, \phi \to N) = \frac{y_D^2 M_N }{8 \pi} e^{-M_N/T_*}$. This calculates the number density of RHNs produced in a Hubble time, after which the Boltzmann suppression becomes more severe as the temperature drops further. There also exist contributions from annihilation processes $\phi \phi \rightarrow N N$ and $\chi\chi\to N N$, but these are subdominant to the inverse decay contribution.

In the presence of a thermal dark sector bath at the time of the FOPT, one also gets a contribution from the up-scattering of $\chi$ states into $N$s due to interactions between the boosted bubble walls and the thermal plasma, with probability \cite{Azatov:2021ifm} 
\beq
\label{eq:partproduction_azatov}
    P(\chi \ra N) \approx \frac{y_D^2 v_{\phi}^2}{M_N^2} \Theta(\gamma_w T_{n} -M_N^2 l_w)\,,
\eeq
which yields the following number density of RHNs 
\beq
\label{eq:Azatovcontribution}
n_N^{wp} \sim  \bigg(\frac{T_{n}}{T_*}\bigg)^3\,n_{th} \, P(\chi \ra N). 
\eeq
In the presence of a thermal bath, it is further possible to produce the RHNs from collisions of high energy particle shells accumulated in front of the expanding bubble walls \cite{Baldes:2023fsp,Baldes:2024wuz}; this contribution is generally subdominant to the above contributions \cite{Giudice:2024tcp}, and we do not include it here. 

There are also several new channels depleting the RHN population. Their effects can be parameterized by $\kappa_{new}$ in \eqref{eq:BA_estimate}. Here we discuss them briefly:

\begin{itemize}
 \item $N \to \chi \phi$. This decay rate into this new channel is approximately
    \beq
    \Gamma_{N \to \chi \phi} \approx \frac{y_{D}^2 M_N}{8 \pi E_N/M_N} \, .
\eeq

    \item $N \, \chi \rightarrow SM$ via s-channel $\phi$ and $N \, SM \to \chi SM $ via t-channel $\phi$ if a thermal bath is present. However, these processes are proportional to the portal coupling $\lambda_{\phi SM }$, and for simplicity we will assume that it is sufficiently small that these processes are negligible.

   \item $N\phi\to SM$ via s-channel $\chi$. This goes through the $\chi$ mixing angle into SM states and is generally suppressed. 
\end{itemize}

Two additional processes are relevant. The first is the additional washout process, $L \, h\to \chi \, \phi$, which can also be incorporated into $\kappa_{new}$. The rate for this process, mediated by the RHN, is similar to that for $L \, h\to \bar{L} \,  h$:
\beq
    \Gamma_{L \, h \rightarrow \chi\phi}\approx n_{th} \frac{y_{\nu}^2\, y_D^2} {8 \pi M_N^2}\,.
\eeq
The analogous process obtained by replacing $\phi$ with its vev, $Lh\to\chi$, is also parametrically similar. The subsequent decay of this light sterile state $\chi$ into SM states can, in principle, also contribute to the lepton asymmetry. However, as mentioned earlier, its contribution is subdominant to that from the RHNs since $m_{\chi} \ll M_N$  in our parameter space, resulting in $\epsilon_{CP}^{\chi}\ll\epsilon_{CP}^{N}$ due to significantly smaller couplings.

\subsection{Parameter Space}

We now explore the parameter space where leptogenesis via bubble collisions can be realized in the neutrino portal setup. 

\begin{figure}[t!]
\centering
\includegraphics[width=0.6\textwidth]{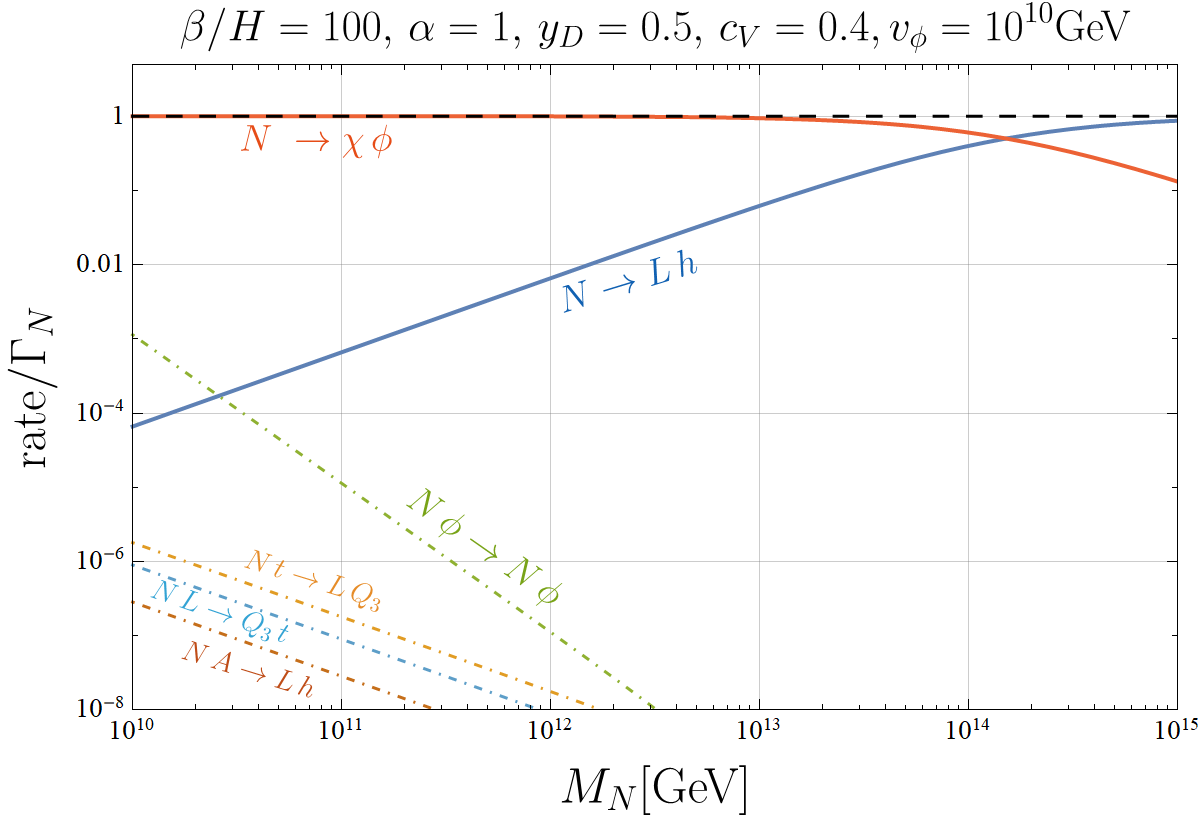}\\
\includegraphics[width=0.6\textwidth]{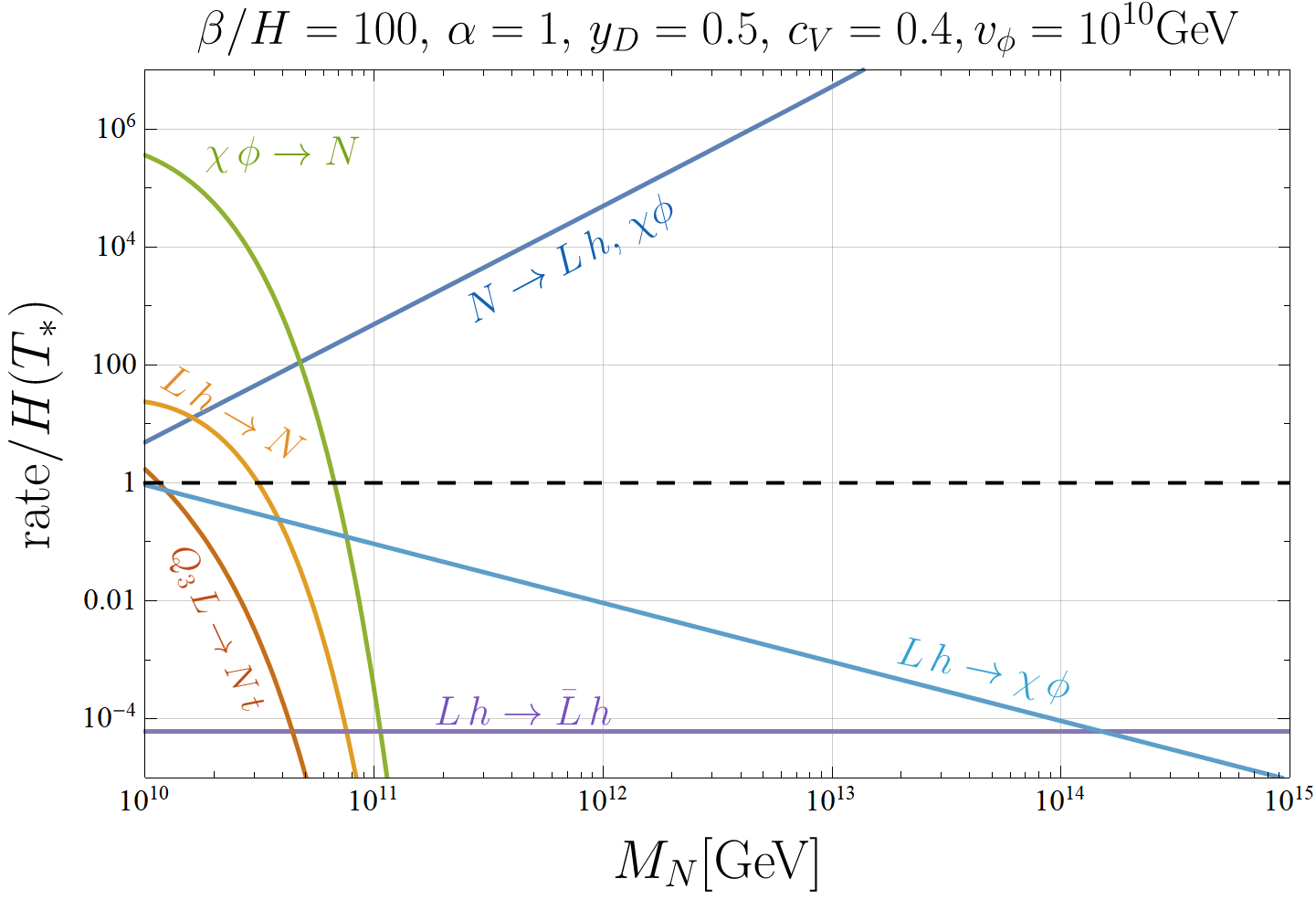}
\caption{Rates for various processes that destroy RHNs (top panel) and wash out the lepton asymmetry (bottom panel). In the top panel, processes that occur only in the presence of a thermal bath are represented by dot-dashed curves.}
\label{fig:leptorates}
\end{figure}

First, we plot in Figure \ref{fig:leptorates} the relative strengths of various processes as a function of the RHN mass for a chosen set of parameters ($\beta/H=100, \alpha=1,y_D=0.5,c_V=0.4, v_\phi=10^{10}$ GeV). In the top panel, we plot the rates for processes that can destroy or modify the RHN abundance within the lifetime of the RHN. The rates are thus normalized to the RHN decay rate $\Gamma_N=\Gamma_{N \to L h}+ \Gamma_{N \to \chi \phi}$. The decay rates into the individual channels $N \to L h $ and $N \to \phi \chi $ are shown in blue and red respectively; since these are proportional to $y_{\nu}^2\approx m_\nu M_N$ and $y_D^2$, their relative contributions to $\Gamma_N$ change accordingly. The remaining curves show processes that can destroy the RHNs or modify their properties through scattering ($N\phi\to N\phi$). The dot-dashed curves correspond to processes that only occur in the presence of a thermal bath; as mentioned earlier, since the thermalization time of the products from bubble collisions is likely significantly longer than the decay lifetime of the RHN, these processes are likely absent if there is no pre-existing thermal bath when the phase transition occurs. Note that the $N\phi\to N\phi$ scattering process also has a nonthermal component, since a population of $\phi$ particles is produced directly from bubble collisions even in the absence of a pre-existing $\phi$ bath; however, this is significantly smaller than the thermal component since the nonthermal number densities from bubble collision are correspondingly small. The main point of observation is that all of these curves for scattering processes are several orders of magnitude below $1$; i.e. all of these processes are unlikely to occur over the lifetime of the RHN, and therefore can be safely neglected in our subsequent calculations.  

In the bottom panel, we plot the rates for various washout processes relative to the Hubble parameter immediately following the phase transition, with a thermal bath at temperature $T_*$. Even if a thermal bath is absent at the time of the transition, these processes get activated as the products of bubble collisions thermalize. Note that all of these processes become weaker at lower temperatures, hence if they are out of equilibrium at this temperature ($T_*$), they will remain out of equilibrium at all subsequent times. The plot shows that when $M_N\sim v_\phi\sim T_*$, most of these processes are active. In particular, $\chi\phi\to N$ (note that this is not a washout process) and $L h\to N $ are rapid and can cause the RHNs to enter thermal equilibrium. The processes that produce an RHN in the final state become inefficient for $M_N/v_\phi>10$ due to the associated kinematic (Boltzmann) suppression. Washout processes that do not carry RHNs in the final state, $Lh\to \bar{L}h, \chi \phi$, drop more slowly at higher RHN masses, as power laws rather than exponentially. Note that $Lh\to \bar{L}h$ remains constant in this plot because the suppression from heavier $M_N$ is compensated by the larger Yukawa couplings $y_\nu$ as enforced by the seesaw relation ($LL\to h h$ scales similarly, hence we do not plot it separately); in contrast, $Lh\to \chi\phi$ drops because we have fixed $y_D$ in this plot, hence some power law suppression with $M_N$ remains. Finally, for comparison, we also plot the (boosted) RHN decay rate relative to the Hubble rate (solid blue curve), which demonstrates that the instantaneous decay approximation for $N$ is fully justified.

\begin{figure}[t!]
\centering
\includegraphics[width=0.59\textwidth]{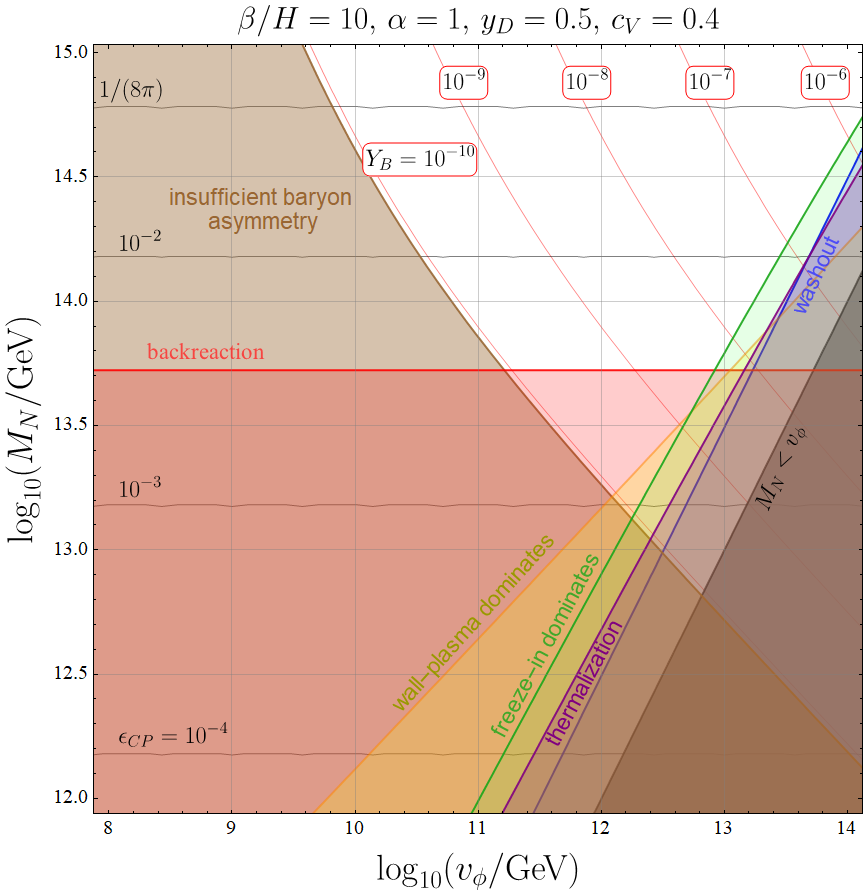}\\
\includegraphics[width=0.59\textwidth]{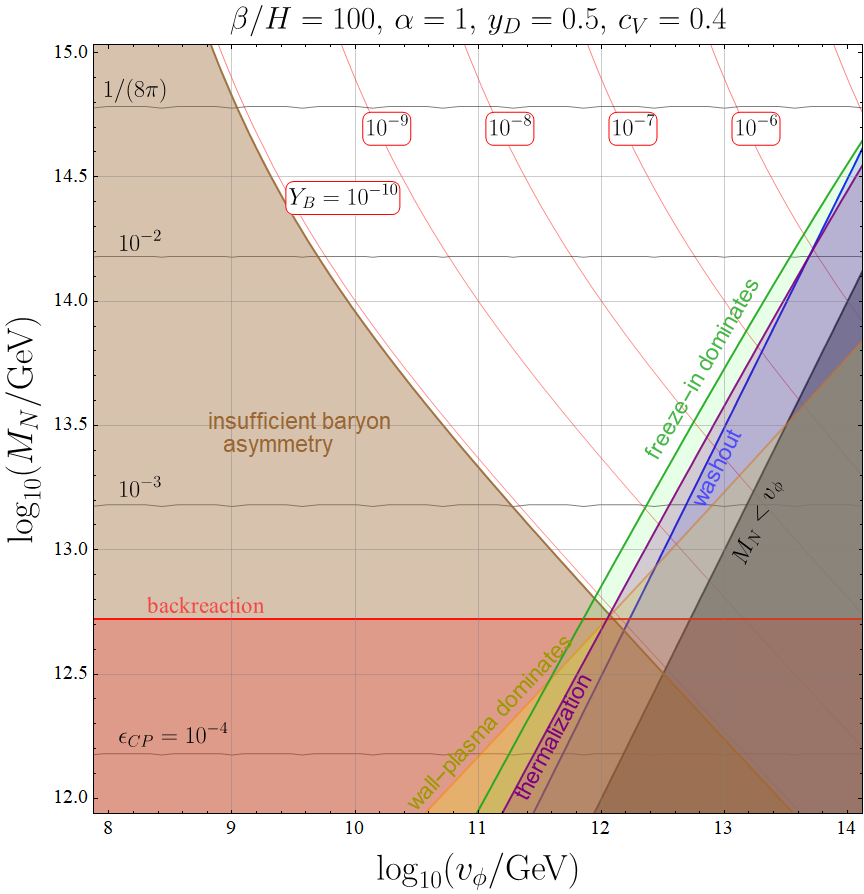}
\caption{Viable parameter space for successful leptogenesis with $\beta /H = 10$ (top panel) and $\beta /H = 100$ (bottom panel). The red contours denote the amount of baryon asymmetry produced ($Y_B$), while the horizontal black contours denote the value of the $CP$-asymmetry parameter $\epsilon_{CP}$. The various colored regions are explained in the text.}
\label{fig:leptoparamspaceneu}
\end{figure}

Next, in Figure \ref{fig:leptoparamspaceneu}, we show the viable parameter space for leptogenesis via bubble collisions in the neutrino portal setup for a set of fixed parameter choices ($\alpha=1, y_D=0.5, c_V=0.4$) and two benchmark values of $\beta /H$ (10 and 100 in the top and bottom panels respectively). In the white region, bubble collisions can produce a sufficient amount of baryon asymmetry; the $Y_B^{bubble}$ contours here represent the estimate from Eq.\,\eqref{eq:BA_estimate}, with $k_{wash} = 1$ and $k_{new} = 1$, i.e.\,assuming washout effects are negligible. For this region, we have explicitly checked that the sphalerons are active when the RHNs decay, so that the produced lepton asymmetry can be efficiently converted into a baryon asymmetry. The various coloured regions represent various constraints:

\begin{itemize}
\item In the brown region, the baryon asymmetry produced is less than the observed value, $Y_B^{bubble}<Y_B^{obs}$. 

 \item In the blue region, washout processes are in equilibrium, i.e. $\Gamma_w > H(T_*)$; the dominant process is either $\Gamma_{L h \to N}$ or $\Gamma_{L h \to \chi \phi}$. However, it is worth noting that even in this region of parameter space, it might still be possible to generate the required amount of baryon asymmetry despite partial washout, but this would require a more careful study, which we do not perform in this paper since our primary focus is on the $M_N\gg v_\phi, T_*$ region. 
 
 \item We also denote regions where RHNs are dominantly produced by other processes that can occur in the presence of a thermal bath. In the yellow region, interactions between the plasma and the bubble wall, as calculated from Eq.\eqref{eq:Azatovcontribution}, create a larger number of RHNs than bubble collisions do. Likewise, the $\chi \phi \to N$  process can dominantly populate the RHNs either via freeze-in (green), as calculated from Eq.\eqref{eq:fi}, or thermalization and freeze-out (purple). In these regions, generation of the required baryon asymmetry remains possible, but the contributions from these additional processes should be taken into account appropriately. 
 
\item In the red region, the energy density in the RHNs exceeds the energy density released from the phase transition, i.e. $n_N^{bubble} E_N > \Delta V (= c_V v_{\phi}^4)$,
where we used $E_N\approx E_{\text{max}}$ as the typical energy of an RHN produced from bubble collision. This implies that backreaction effects of particle production on the bubble collision dynamics become important and should be taken into account, hence the above calculations are not valid in this regime.
\item  In the gray region, $M_N< v_\phi$, and several assumptions in the particle production calculation formalism break down, hence the calculation is no longer reliable (see 
\cite{Giudice:2024tcp} for more detailed discussions).  
\end{itemize}

We have also checked that scattering processes remain subdominant to the decay processes for $M_N>v_\phi$. More generally, we see all scattering and washout processes can be neglected for $M_N/v_\phi\gtrsim 10$. Comparing the two panels, we see that a shorter duration of the phase transition, i.e. larger value of $\beta/H$, opens up more paramater space for leptogenesis, as a larger $\beta/H$ implies a greater number of bubbles per Hubble volume, hence more collisions and greater particle production. At the same time, it also has the beneficial effect of reducing the backreaction regime, as the typical RHN energy (given by $E_{\text{max}}$) is lower, hence the RHNs take away less energy.   

\begin{figure}[t!]
\centering
\!\!\!\!\!{\includegraphics[width=0.49\textwidth]{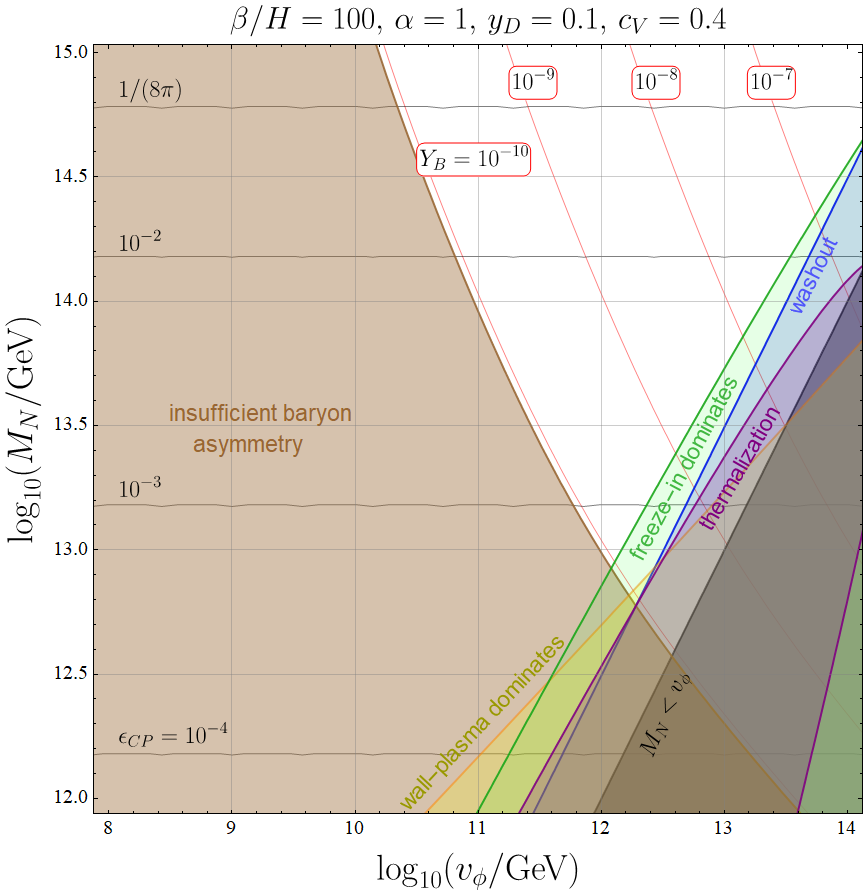}}
\hspace{1mm}
{\includegraphics[width=0.49\textwidth]{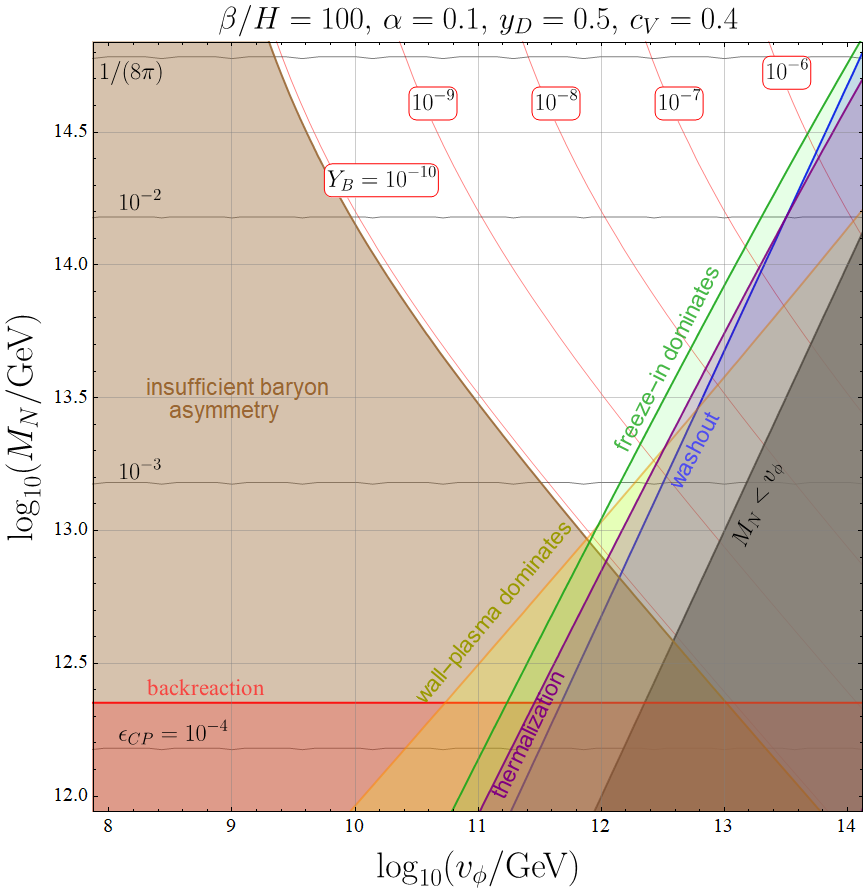}}\\
\!\!\!\!\!{\includegraphics[width=0.49\textwidth]{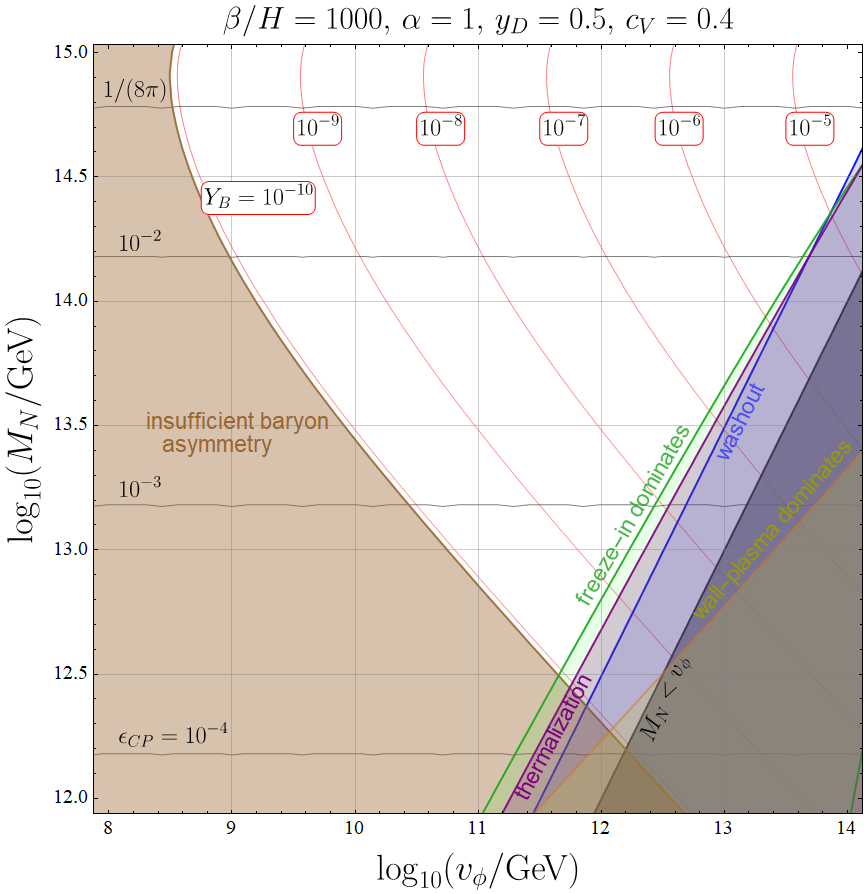}}
\hspace{1mm}
{\includegraphics[width=0.49\textwidth]{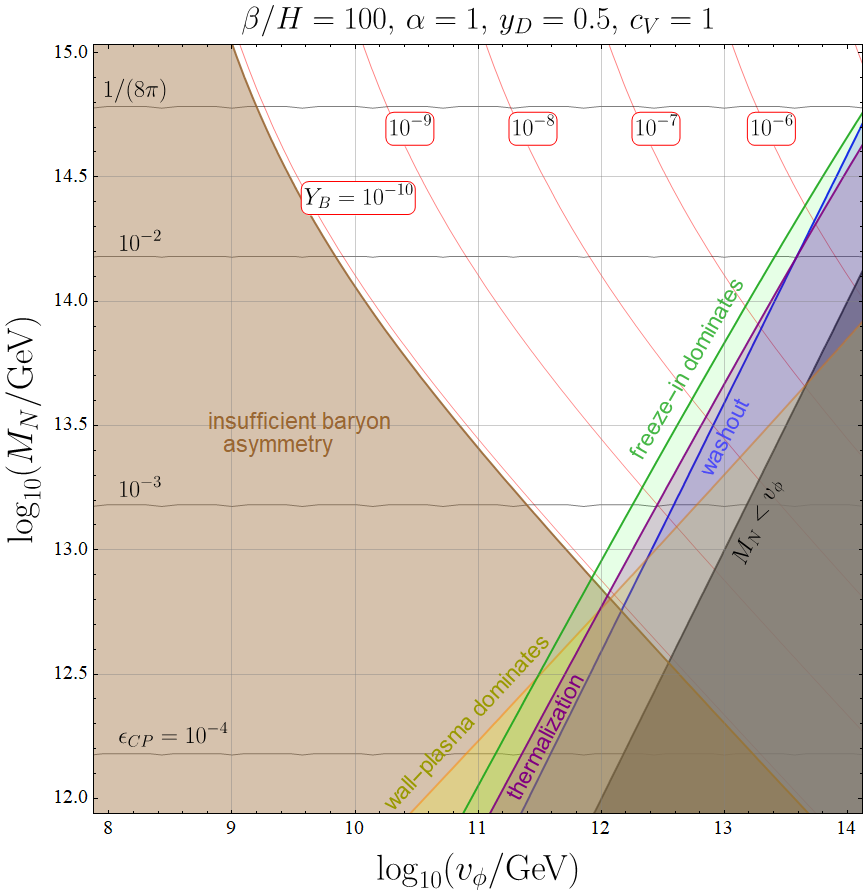}}
\caption{Modified parameter space from changing the following parameters relative to the bottom panel of Fig.\,\ref{fig:leptoparamspaceneu}: $y_D=0.1$ (top left panel), $\alpha=0.1$ (top right), $\beta/H=1000$ (bottom left), and $c_V=1$ (bottom right).
}
\label{fig:leptoparamspaceneu1000}
\end{figure}

Figure \ref{fig:leptoparamspaceneu1000} shows how the parameter space evolves relative to the bottom plot of Fig.\,\ref{fig:leptoparamspaceneu} when some of the parameters are changed. In the top left panel, we changed $y_D=0.1$, which results in a suppression of the number density of RHNs produced from the bubble collisions, hence a reduction in the produced lepton asymmetry, as well as the elimination of the backreaction region. However, note that all processes involving the RHNs and dark sector particles, including washout processes, are also similarly suppressed. The top right panel features a lower value of the strength of the phase transition, $\alpha=0.1$. This represents a smaller amount of latent heat available in the transition, hence a smaller number of RHNs produced from bubble collisions. Meanwhile, processes associated with the thermal bath (e.g. freeze-in, washout) remain unchanged, hence we see that they become relevant in a larger region of parameter space. In the bottom left panel we set $\beta/H =1000$; this implies a shorter phase transition, hence smaller bubbles and a greater number of collisions, resulting in enhanced production of RHNs, opening up the parameter space available for leptogenesis. Finally, the bottom right panel features $c_V=1$. Since $\alpha$ is held fixed, this results in an increase in the reheat tempetaure $T_*$ for a given phase transition scale $v_\phi$, which ultimately decreases the RHN yield produced from bubble collisions (as can be inferred from Eq.\,\ref{eq:neu_Y_neu}) and reduces the viable parameter space.

\begin{figure}[t!]
\centering
\includegraphics[width=0.6\textwidth]{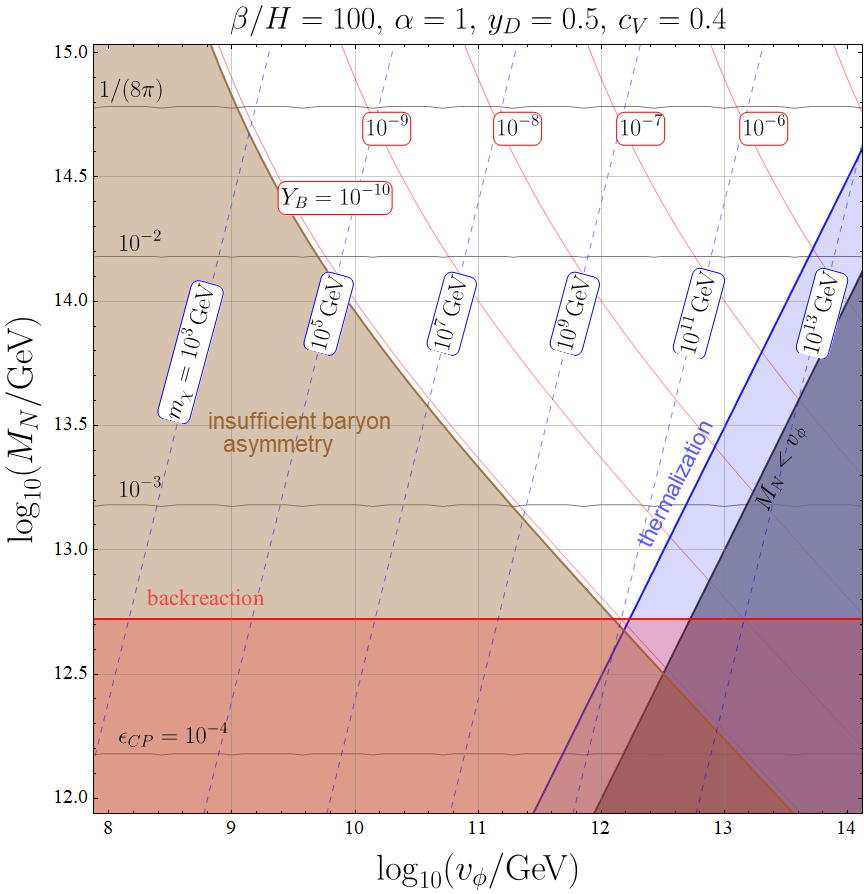}\\
\caption{Contours of $m_\chi$ (blue dashed lines) overlaid on the leptogenesis parameter space from the bottom panel of Fig.\,\ref{fig:leptoparamspaceneu}.}
\label{fig:mchi}
\end{figure}

In Figure \ref{fig:mchi}, we present some properties of the light dark sector fermion $\chi$ overlaid on the parameter space from the bottom panel of Fig.\,\ref{fig:leptoparamspaceneu}. The blue contours denote the mass of $\chi$; as anticipated before, since it obtains its mass from the seesaw mechanism in the dark sector, it is generally significantly lighter than the scale of symmetry breaking if $M_N\gg v_\phi$. In the region of parameter space of interest to us, its mass is at or above the TeV scale, and we have checked that its decay occurs before Big Bang Nucleosynthesis (BBN) and therefore is not problematic for cosmology. As previously mentioned, the $CP$-asymmetry associated with $\chi$ decay is significantly smaller due to the suppressed effective coupling between $\chi$ and the SM neutrinos, hence they contribute negligibly to the generation of the lepton asymmetry. In the blue region, $\chi$  thermalizes with the SM bath via the process $L \, h \to \chi \phi$ (this also corresponds to the region from Fig.\,\ref{fig:leptoparamspaceneu} where washout effects are efficient).

\begin{figure}[t!]
\centering
\includegraphics[width=0.7\textwidth]{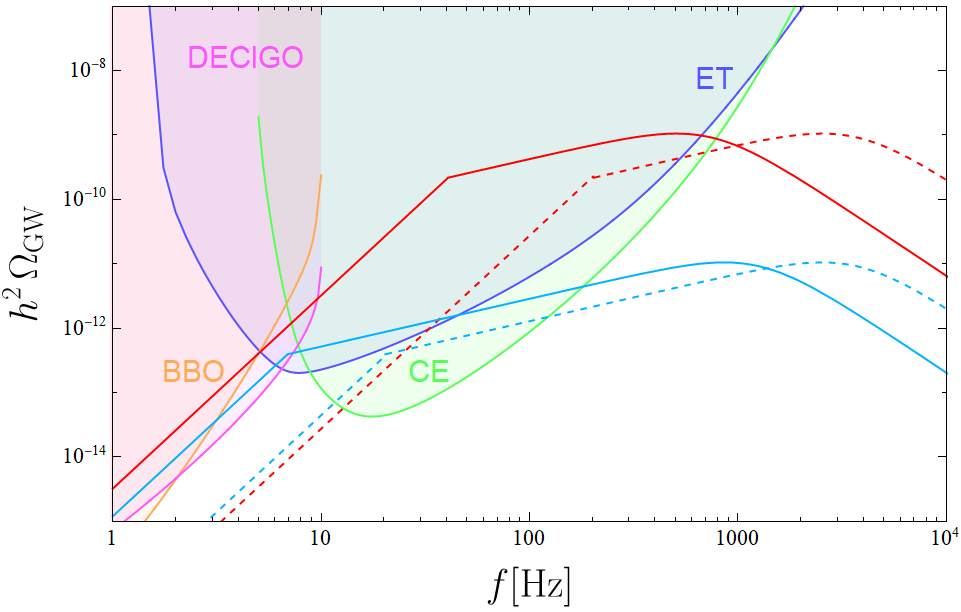}\\
\caption{Gravitational wave signals for some benchmark cases: $\beta / H = 10$ with  $v_{\phi} \simeq 4 \times 10^{9}$ GeV (red solid curve) and $v_{\phi} \simeq 2 \times 10^{10}$ GeV (red dashed curve), $\beta / H = 100$ with $v_{\phi} \simeq 7 \times 10^{8}$ GeV (light blue solid curve) and $v_{\phi} \simeq 2 \times 10^{9}$ GeV (light blue dashed curve). We set $\alpha = 1$, $c_V=0.4$. For reference, we plot the power law integrated sensitivity curves for the Einstein Telescope (ET) \cite{Punturo:2010zz}, Cosmic Explorer (CE) \cite{Reitze:2019iox}, Big Bang Observer (BBO) \cite{Harry:2006fi}, and Deci-hertz Interferometer Gravitational wave Observatory (DECIGO) \cite{Kawamura:2006up}, corresponding to one year observation time, with signal-to-noise ratio $=1$, obtained from \cite{Schmitz:2020syl}.}
\label{fig:GWsignal}
\end{figure}

In Figure \ref{fig:GWsignal}, we plot the gravitational wave signals (calculated using the formulae in Sec.\,\ref{subsec:gw}) expected from some benchmark FOPTs that can produce the required amount of lepton asymmetry against the reach of several planned future GW experiments (see figure caption for details). We therefore expect GW signals at future GW experiments to be an observable 
aspect of leptogenesis via bubble collisions.  

\section{Scalar Portal}
\label{sec:splepto}

In this section, we consider a qualitatively different realization of RHN production from bubble collision. In contrast to the neutrino portal setup discussed in the previous section, consider instead the case where there is no direct coupling between $\phi$ and $N$, and a heavy singlet scalar field $S$ instead acts as the portal between the two fields. This scalar field could, for instance, be responsible for the spontaneous breaking of lepton number at some very high scale, and be correspondingly heavy. $S$ can then couple to $\phi$ through a scalar portal coupling without being charged under the symmetry broken by $v_\phi$.

\subsection{Setup}

The new field content and interactions in this case are given by
\beq
\mathcal{L}\supset \frac{\lambda_S}{4}\,|\phi|^2 |S|^2+m_S^2 S^2+y_\nu \overline{L} h N+ y_S S N N+M_N \overline{N^c}N\,.
\label{splag}
\eeq
Note that the coexistence of the final two terms implies that $S$ obtains a vev $\langle S \rangle$, and it is then plausible that $M_N=y_S \langle S \rangle$, as would be the case if $\langle S\rangle$ breaks lepton number, although we will treat $y_S$ and $M_N$ to be independent parameters in our study for generality. We will further assume $m_S> 2 M_N$, so that the singlet can decay into a pair of RHNs, $S\to NN$. 

Compared to the neutrino portal setup described above, the scalar portal setup has the advantage that there are no light fields ($\chi$) that could be problematic for cosmology. The additional field content instead is the heavy scalar field $S$, which couples the FOPT background field $\phi$ to the RHN. The disadvantage of this setup is that $\phi$ does not couple directly to $N$, hence the production of RHNs from bubble collisions will be relatively inefficient. 

\subsection{RHN Abundance}

In this setup, the RHNs are not produced directly from bubble collisions. Bubble collisions instead produce the scalar $S$ via $\phi^*\to \phi S$ (obtained from inserting the singlet vev $\langle S \rangle$ into the first term in the Langrangian above). There also exists another two-body  decay channel $\phi^*\to S S$, which is generally subdominant, as well as the three body decay $\phi^*\to \phi S S$, which can dominate if $E_{\text{max}}\gg \langle S \rangle$.

The RHNs are produced from the decay of the scalar, $S\to NN$. The RHN number density in this case can be written as 
\beq
\label{eq:neu_numbdens_scalar}
n_N\approx 5.2\times 10^{-3}\frac{\beta}{H}\left(\frac{30(1+\alpha)c_V}{\pi^2\alpha}\right)^{1/2} \frac{v_\phi^4}{M_{Pl}}\frac{\lambda_S^2 y_S^2}{\lambda_S^2+y_S^2}\left[\text{ln}\left(\frac{E_{\text{max}}}{m_S}\right)+\frac{8\pi^2 \langle S \rangle^2}{m_S^2}\right] \, ,
\eeq
with $E_{\text{max}}$ as given in Eq.\,\ref{emax}. The two terms in the square parenthesis denote contributions from $\phi^*\to \phi S S$ and $\phi^*\to \phi S$ respectively. Here we will assume $\langle S \rangle\approx m_S$. Since the logarithmic factor is $\mathcal{O}(1)$, the second term tends to dominate the number density. In this case, the $S$ particles are produced essentially at rest, hence each $N$ has typical energy $m_S/2$ (note that this is in contrast to the neutrino portal setup, where the RHNs had typical energy $E_N\sim E_{\text{max}}$), and the RHN decay width is $\Gamma_D = \frac{y_{\nu}^2 M_N}{8 \pi} \frac{M_N}{m_s/2}$ . 

Using the above expression, the RHN yield is
\beq
\label{eq:neu_Y_scalar}
Y_N=\frac{n_N}{s}\approx \frac{1}{8 \pi^2} \frac{\lambda_S^2 y_S^2}{\lambda_S^2+y_S^2}\, \frac{\beta}{H}\left(\frac{\pi^2\alpha}{30(1+\alpha)g_* c_V}\right)^{1/4} \frac{v_\phi}{M_{Pl}}\,\left[\text{ln}\left(\frac{E_{\text{max}}}{m_S}\right)+\frac{8\pi^2 \langle S \rangle^2}{m_S^2}\right]\,.
\eeq

In principle, by integrating out the scalar $S$, we can also obtain direct couplings between $\phi$ and the RHNs, enabling the decays $\phi^*\to NN, \phi NN$. However, these are suppressed relative to the cascade decay chain above due to the additional couplings and the $S$ propagator, and we have explicitly checked that it never dominates in the region of parameter space of interest to us. We will therefore ignore this contribution in our study. 

In addition to bubble collisions, there exist other mechanisms that can produce the scalar $S$. The process $\phi\phi\to S$ can thermalize it with the bath, or produce a freeze-in abundance
\beq
\label{eq:fi2}
    n_{S}^{FI}= \frac{\Gamma(\phi \, \phi \to S )}{H(T_*)}n_{th} 
\eeq
with $\Gamma(\phi \, \phi \to S ) = \frac{\lambda_S^2 \langle S \rangle^2 }{8 \pi m_S}e^{- m_S/T_*}$. 
If the bubbles expand into a thermal dark sector bath, $\phi$ can also upscatter into $S$ with  probability 
\beq
\label{eq:partproduction_azatov}
    P(\phi \ra S) \approx  \frac{\lambda_S^2 v_{\phi}^2  \langle S \rangle^2}{ m_S^4} \Theta(\gamma_w T_{nuc} -  m_S^2 l_w) \, .
\eeq
Thus, 
\beq
\label{eq:Azatovcontributionscal}
n_S^{wp} =  n_{th} P(\phi \ra S) \bigg(\frac{T_{n}}{T_*}\bigg)^3 \,.
\eeq

In this scalar portal setup, there are several new processes involving $S$ and $N$ to take into account:

\begin{itemize}
\item $S  \to \phi \phi $. This is an additional decay channel for $S$, obtained by inserting $\langle S\rangle$ in the first term in Eq.\,\ref{splag}, with decay rate $\Gamma_{S \ra \phi \phi} =  \frac{\lambda_S^2 \langle S \rangle^2 }{8 \pi m_S}$. This suppresses the production of RHNs by the branching ratio $Br(S \to N N) = \Gamma_{S \ra N N}/(\Gamma_{S \ra N N} + \Gamma_{S \ra \phi \phi})$.

 \item $S \, S \to \phi \phi $. Its interaction rate is $
      \Gamma_{S S \ra \phi \phi} = n_{S}^{bubble} \frac{\lambda_S^2}{ 16 m_S^2} \frac{1}{4 \pi^2}
      $. We have checked that this process is subdominat to $S \to \phi \phi$, thus we neglect it. $SS\to NN$ is similarly subdominant to $S\to NN$.      
    \item $N \, N \to S$. Due to the small number density of $N$s produced from bubble collisions, this process is generally negligible.
\end{itemize}

\subsection{Parameter Space}

\begin{figure}[t!]
\centering
\includegraphics[width=0.6\textwidth]{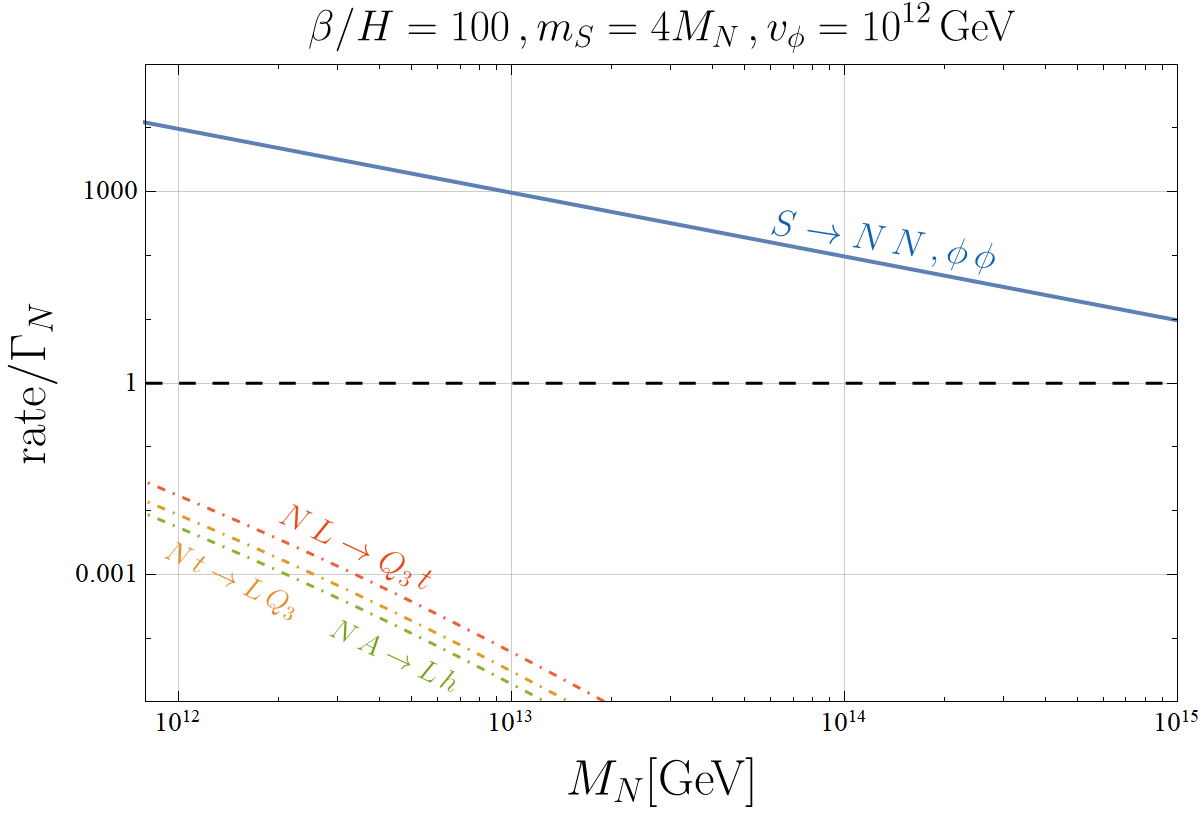}\\
\includegraphics[width=0.6\textwidth]{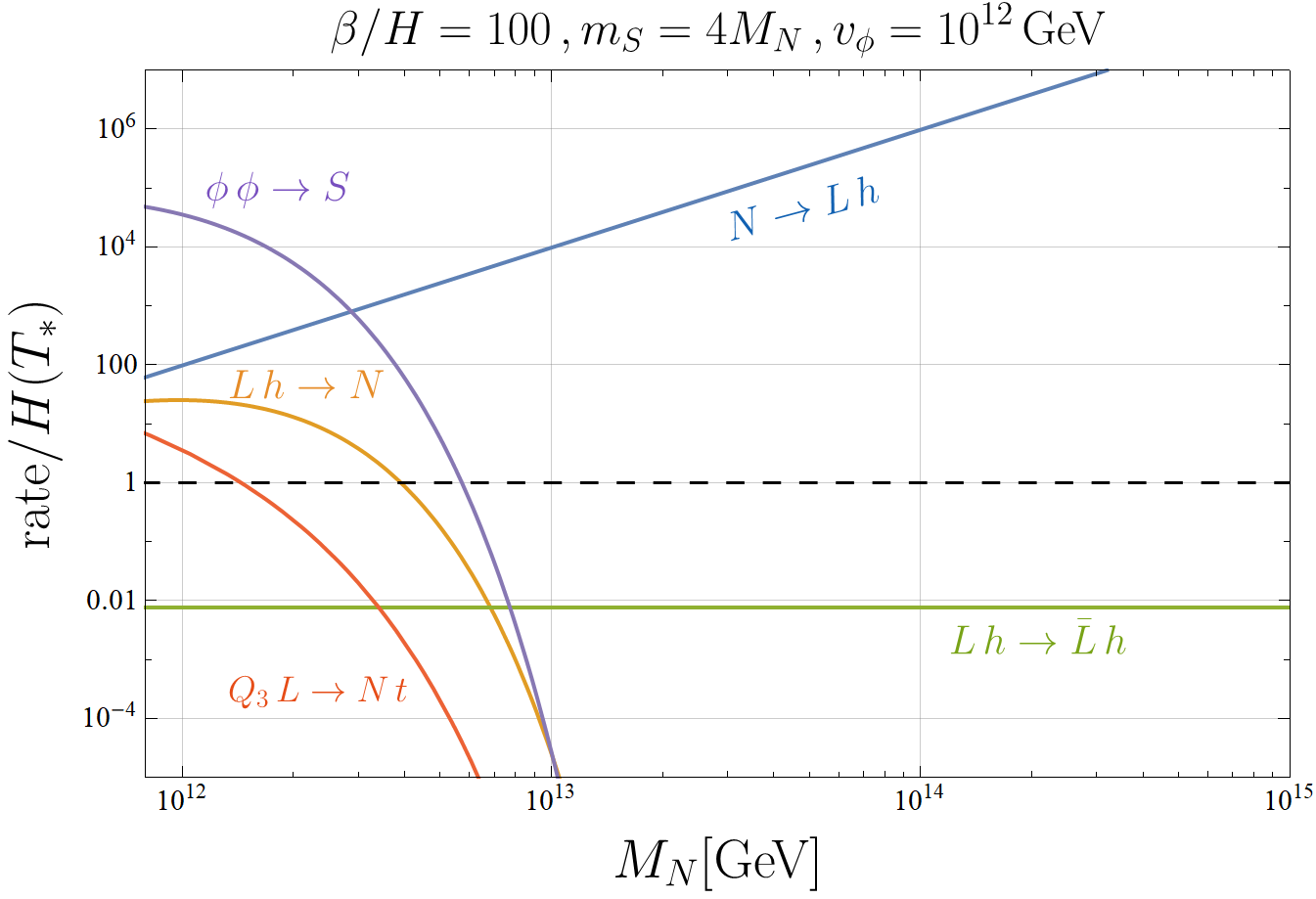}
\caption{Rates for various processes that destroy RHNs (top panel) and wash out the lepton asymmetry (bottom panel). We have set $\alpha=c_V=\lambda_S=1$. In the top panel, processes that occur only in the presence of a thermal bath are represented by dot-dashed curves. }
\label{fig:scalarrates}
\end{figure}

In Figure \ref{fig:scalarrates}, we plot the relative rates for various processes in the scalar portal setup, analogous to Fig.\,\ref{fig:leptorates} above for the neutrino portal setup. For this plot, we set $\beta/H = 100$, $m_S=4 M_N$, $v_{\phi}= 10^{12}$ GeV, and $\alpha = c_V = \lambda_S=1$. In the top panel, the rates of processes destroying/modifying the RHN population are normalized by the decay width $\Gamma_N = \Gamma_{N \to L h}$, where the RHNs are now boosted with energy $E_N \approx m_S/2$, instead of $E_{\text{max}}$. Thus, the rates of scatterings become more relevant here compared to the neutrino portal case, but still remain negligible. For comparison, we also plot the decay rate of the scalar $S$. 
In the bottom panel, we plot the rates of various washout processes normalized to the Hubble rate ($\phi\phi \to S$ is not a washout process, but we include it here for reference). As was the case for the neutrino portal setup, we see that the washout processes are out of equilibrium and can be neglected for $M_N/v_{\phi} \gtrsim 10$.

\begin{figure}[t!]
\centering
\!\!\!\!\!{\includegraphics[width=0.49\textwidth]{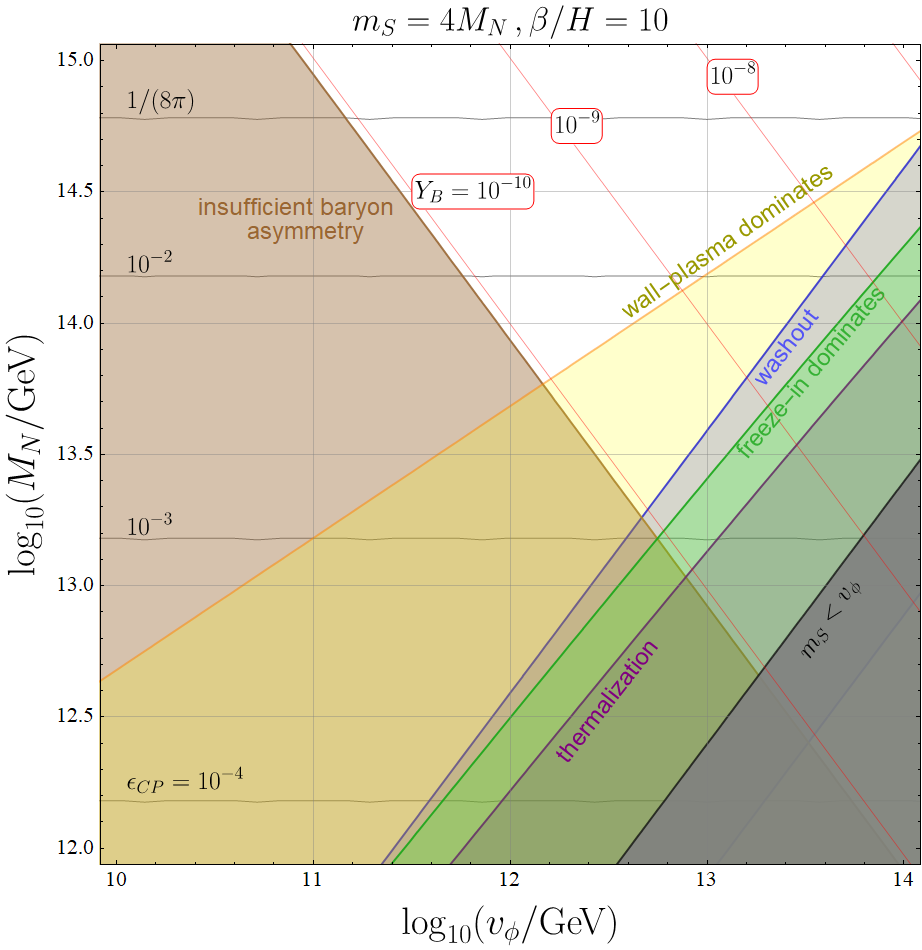}}
\hspace{1mm}
{\includegraphics[width=0.49\textwidth]{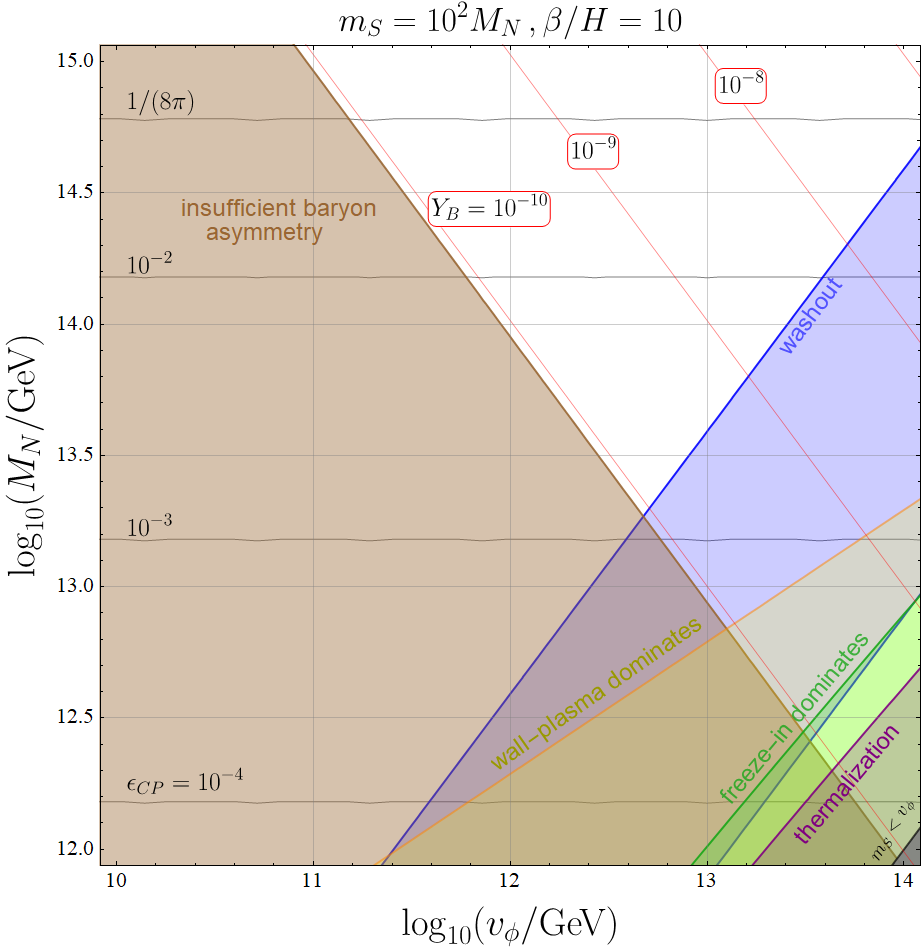}}\\
\!\!\!\!\!{\includegraphics[width=0.49\textwidth]{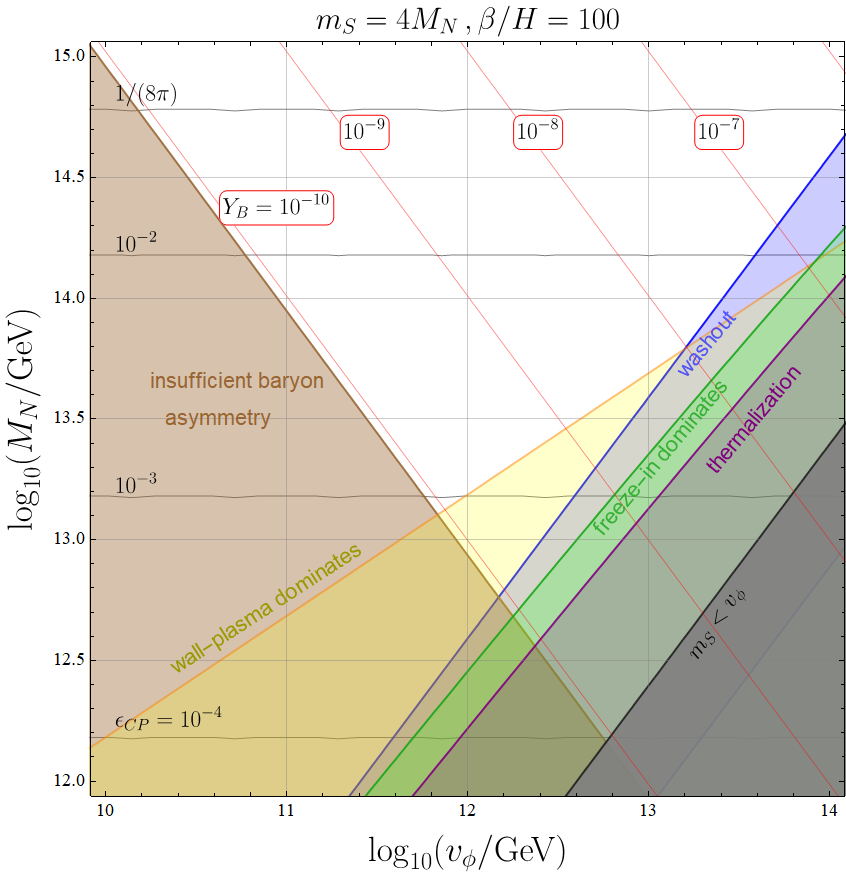}}
\hspace{1mm}
{\includegraphics[width=0.49\textwidth]{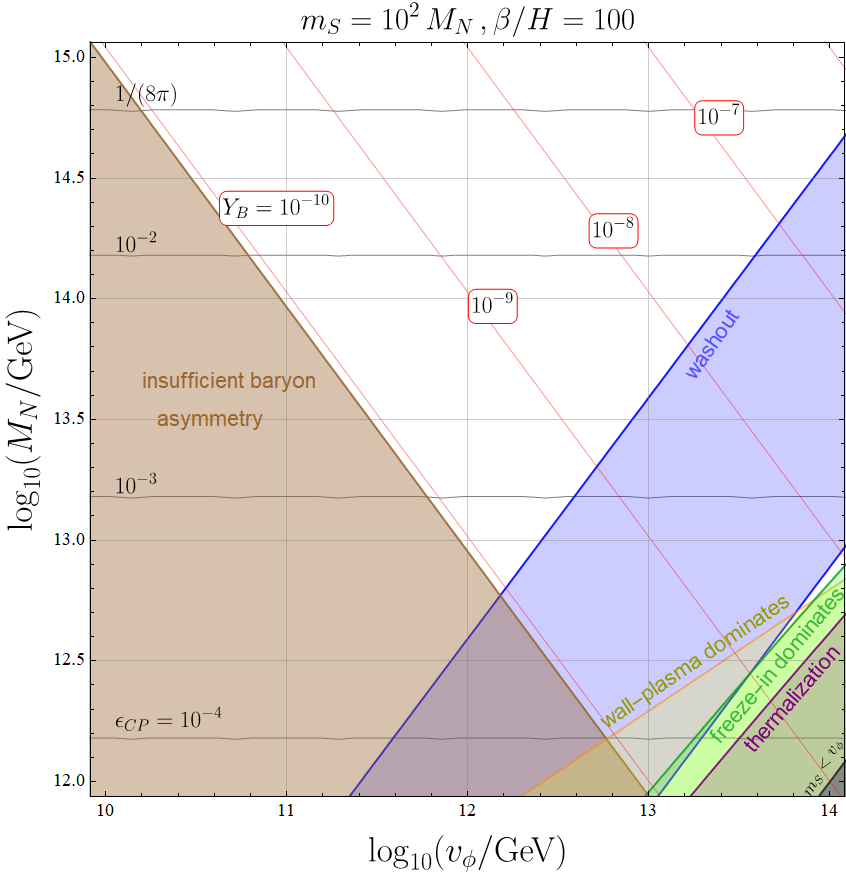}}
\caption{Viable parameter space for successful leptogenesis in the scalar portal setup, for various choices of $\beta/ H$ and $m_S$. We set $\alpha=c_V=\lambda_S=1$.
See text for details.}
\label{fig:paramspacescal1000}
\end{figure}

Figure \ref{fig:paramspacescal1000} shows the relevant parameter space for  successfully producing the baryon asymmetry in the scalar portal setup. We show two benchmark cases for the scalar mass, $m_S=4M_N$ in the left plots and $m_S=10^2 M_N$ in the right plots, and two choices for the duration of the phase transition, $\beta / H = 10$ and $\beta / H = 100$ in the top and bottom rows, respectively. The various contours and colored regions carry the same meanings as the neutrino portal plots (Fig.\,\ref{fig:leptoparamspaceneu}), see text in the previous section for details. 

From Eq.\,\ref{eq:neu_numbdens_scalar}, we see that increasing $m_S$ does not modify the number density of RHNs (assuming $\langle S \rangle = m_S$), hence the brown regions do not shift between the left and right panels. However, increasing $m_S$ makes $S$ production from thermal processes inefficient, hence significantly reducing the wall-plasma (yellow) and freeze-in (green) contributions. Increasing $\beta/H$ increases the number of bubble collisions and hence the RHN abundance, thereby opening up more parameter space, as in the neutrino portal setup. It is worth noting that there is no regime where backreaction is relevant in these plots (compare with the corresponding plots for the neutrino portal setup, Fig.\,\ref{fig:leptoparamspaceneu}). This is because the RHNs are now no longer boosted with energy $\sim E_{\text{max}}$, but instead carry away significantly less energy $E_N\approx m_S/2$. We see that washout effects become irrelevant for $M_N/v_\phi\gtrsim 10$, as in the neutrino portal case. Overall, the scalar portal setup is less efficient than the neutrino portal setup in producing the lepton asymmetry, as bubble collisions have to produce the heavier scalar $S$ states rather a fermion pair containing the RHN (as in the neutrino portal case), which is a less efficient process.

\section{Summary}
\label{sec:summary}

Here we summarize the key results of our paper

\begin{itemize}
\item We have demonstrated that collisions of runaway bubbles at a first order phase transition (FOPT) provide a viable nonthermal mechanism for the production of heavy sterile neutrinos and leptogenesis. We have demonstrated the viability of this mechanism in two qualitatively different setups: the neutrino portal setup (Section \ref{sec:nplepto}), where the RHNs couple directly to the field undergoing the phase transition, and the scalar portal setup (Section \ref{sec:splepto}), where a heavy scalar provides the portal between these fields. 

\item This mechanism can produce RHNs with mass several orders of magnitude beyond the scale of symmetry breaking and the temperature of the plasma. Thus, there is no need for a reheat temperature above the RHN mass scale, as is the case for standard thermal leptogenesis. Due to this separation of scales, all washout processes arising from the thermal bath are naturally suppressed. Our results indicate that this is true for $M_N/v_\phi>10$.

\item Leptogenesis with RHN masses $\gtrsim 10^{14}$ GeV, the natural scale for type-I seesaw with $\mathcal{O}(1)$ couplings, is readily realized. Note that standard thermal leptogenesis with such masses faces strong suppression from washout processes that are in equilibrium at such temperatures. 

\item Gravitational waves provide a testable signal of this mechanism. We find that leptogenesis through bubble collisions can occur for phase transitions at scales $\gtrsim 10^9$ GeV. For sufficiently strong phase transitions ($\alpha\sim 1$), the associated gravitational wave signals can be observable with next generation telescopes such as the Einstein Telescope and Cosmic Explorer, and with next generation space-based interferometers such as Big Bang Observer and Deci-hertz Interferometer Gravitational wave Observatory (see Figure \ref{fig:GWsignal}).

\end{itemize}

\section*{Acknowledgements}
We are grateful to Simone Blasi, Geraldine Servant, and Thomas Konstandin for helpful discussions. The authors are supported by the Deutsche Forschungsgemeinschaft under Germany's Excellence Strategy - EXC 2121 ``Quantum Universe" - 390833306. B.S. also thanks New York University, Johns Hopkins University, and the Munich Institute for Astro-, Particle and BioPhysics (MIAPbP), where parts of this project were completed, for hospitality.

\appendix

\bibliographystyle{JHEP}
{\footnotesize
\bibliography{LeptogenesisBubbleCollisions}

\end{document}